\documentclass[letterpaper, 11pt]{article}
\usepackage{amsmath,amsthm,graphicx,url}
\usepackage{amssymb}
\usepackage[margin=1in]{geometry}
\usepackage{booktabs} 
\usepackage[ruled,noline]{algorithm2e} 
 
 \SetKwBlock{Initialize}{Initialize:}{} \SetKwFor{From}{from}{do}{}
\SetKwFor{ForAllInParallel}{for all}{in parallel do}{}
\SetAlFnt{\small}
\SetAlCapFnt{\small}
\SetAlCapNameFnt{\small}
\SetAlCapHSkip{0pt}
\IncMargin{-\parindent}
\usepackage{authblk}
\newcommand*\samethanks[1][\value{footnote}]{\footnotemark[#1]}
\usepackage[numbers]{natbib}

 \usepackage{amsmath,amsthm,graphicx,url}
\usepackage{tikz}
\usepackage{subcaption}
 \usepackage{mathtools}
\usepackage[normalem]{ulem}
\usepackage{soul}
\usepackage{graphicx,subcaption}

 \usepackage[ruled, noline]{algorithm2e} 

\usepackage[]{color-edits}
\addauthor{VG}{red}

\addauthor{MF}{blue}

\addauthor{BB}{magenta}

\addauthor{XT}{cyan}

\newcommand{\R}{\mathbb{R}}
\newcommand{\V}{\mathcal{V}}                                
\newcommand{\C}{\mathcal{C}}   
\newcommand{\X}{\mathcal{X}}                                
\renewcommand{\d}{d}                                        
\newcommand{\pref}{\succcurlyeq}                            
\newcommand{\SC}{\mathsf{SC}}                               
\newcommand{\dist}{\mathsf{distortion}}                           
\newcommand{\alg}{\mathsf{ALG}}                             
\newcommand{\SimVeto}{\textsc{LA}_\delta}     
\newcommand{\p}{\hat{c}}                                          
\renewcommand{\o}{c^*}              

\newcommand{\error}{\eta}

\newcommand{\topc}{\texttt{top}}
\newcommand{\plu}{\texttt{plu}}

\newcommand{\m}{\mu}
\newcommand{\unip}{p^\texttt{uni}}
\newcommand{\vdg}{\texttt{VDG}}
  \theoremstyle{definition}
\newtheorem{theorem}{Theorem}[section]

 \newtheorem{observation}[theorem]{Observation}
\usepackage{thm-restate}

\theoremstyle{acmplain}

\newtheorem{definition}[theorem]{Definition}
 \newtheorem{proposition}[theorem]{Proposition}
 \newtheorem{lemma}[theorem]{Lemma}
 \newtheorem{corollary}[theorem]{Corollary}

\title{Learning-Augmented Metric Distortion via $(p,q)$-Veto Core} 

\author[a]{Ben Berger}
\author[b]{Michal Feldman\thanks{Partially supported by the European Research Council (ERC) under the European Union's Horizon 2020 research and innovation program (grant agreement No. 866132), by an Amazon Research Award, and by the NSF-BSF (grant number 2020788).}}
\author[c]{Vasilis Gkatzelis\thanks{Partially supported by NSF CAREER award CCF-2047907 and NSF grant CCF-2210502.}}
\author[c]{Xizhi Tan\samethanks}

\affil[a]{Offchain Labs, inc.;
\texttt{bberger@offchainlabs.com}
}

\affil[b]{Tel-Aviv University and Microsoft ILDC; \texttt{{ mfeldman@cs.tau.ac.il}}
}

\affil[c]{Drexel University;
\texttt{{\{gkatz,xizhi\}@drexel.edu}}}

\date{}
\begin{document}

\begin{titlepage}

\maketitle
\begin{abstract}
In the metric distortion problem there is a set of candidates $C$ and voters $V$ within the same metric space. The goal is to select a candidate minimizing the social cost, defined as the sum of distances of the selected candidate from all the voters, and the challenge arises from the algorithm receiving only \emph{ordinal} input --- each voter's list of candidates ranked by distance --- while the objective function is \emph{cardinal}, determined by the underlying metric. The \emph{distortion} of an algorithm is its worst-case approximation factor with respect to the optimal social cost.

A key concept here is the $(p,q)$-\emph{veto core}, with $p\in \Delta(V)$ and $q\in \Delta(C)$ being normalized weight vectors representing voters' veto power and candidates' support, respectively. The $(p,q)$-veto core corresponds to a set of winners from a specific class of deterministic algorithms. Notably, the optimal distortion of $3$ is obtained from this class, by selecting veto core candidates using uniform $p$ and $q$ proportional to candidates' plurality scores. Bounding the distortion of other algorithms from this class 
is an
open problem.

Our contribution is twofold. First, we establish upper bounds on the distortion of candidates from the $(p,q)$-veto core for arbitrary weight vectors $p$ and $q$. Second, we revisit the metric distortion problem through the \emph{learning-augmented} framework, which equips the algorithm with a (machine-learned) prediction regarding the optimal candidate. The quality of this prediction is unknown, and the goal is to optimize the algorithm's performance under accurate predictions (known as \emph{consistency}), while simultaneously providing worst-case guarantees under arbitrarily inaccurate predictions (known as \emph{robustness}). We propose an algorithm that chooses candidates from the $(p,q)$-veto core, using a prediction-guided $q$ vector and, leveraging our distortion bounds, we prove that this algorithm achieves the optimal robustness-consistency trade-off. 
\end{abstract}

\thispagestyle{empty} 
\end{titlepage}

\setcounter{page}{1} 
\section{Introduction}
\label{sec:introduction}
In the metric distortion problem there is a set of candidates $\C$ and a set of voters $\V$, all residing in the same metric space. This metric defines a distance $d(v,c)$ between each voter $v\in\V$ and candidate $c\in \C$, and the objective is to choose a candidate $c$ with minimum social cost, defined as the total distance of the chosen candidate from all voters, {\sl i.e.}, $\sum_{v\in \V} d(v,c)$. This captures spatial models of voting from the political science literature, such as the Downsian proximity model~\cite{enelow1984spatial}, as well as the classic problem of facility location~\cite{charikar1999guha,JMMSV03,cohen2019polynomial}. Computing the optimal candidate, $\o=\arg\min_{c\in \C}\sum_{v\in \V}d(v,c)$, given the pairwise distances, $d(v,c)$, is a computationally easy problem, but the challenge in the metric distortion problem is that the algorithm has access only to {\em ordinal} information regarding the preferences of each voter, {\sl i.e.}, a ranked list of candidates in non-decreasing order of their distance from the voter. This information limitation restricts the designer's ability to optimize the social cost and the performance of an algorithm is evaluated based on its {\em distortion}, {\sl i.e.}, its worst-case approximation factor with respect to the optimal social cost. 

The metric distortion problem has been the focus of a series of papers
\cite{
anshelevich2018metric,
munagala2019improved,
GHS20,
KK22,
KK23,
kempe2020duality,
peters2023note} which evaluated
the distortion of known algorithms and designed new ones, aiming to achieve better distortion bounds. \citet{anshelevich2018metric} established that no deterministic algorithm can achieve a distortion better than 3, and \citet{GHS20} eventually provided a
deterministic algorithm that matches this bound (see Section~\ref{sec:related_work} for details). Even after achieving this tight bound, subsequent work by~\citet{KK22,KK23} and \citet{peters2023note} has deepened our understanding of the metric distortion problem, leading to the formulation of simpler algorithms that achieve the same bound.

A key concept resulting from this line of work is the $(p,q)$-\emph{veto core}, where $p\in \Delta(V)$ and $q\in \Delta(C)$ are normalized weight vectors representing the voters' veto power and candidates' support, respectively. \citet{KK23} showed that the $(p,q)$-veto core corresponds to a set of candidates that arise as winners under a natural class of algorithms. Notably, all known algorithms that achieve the optimal distortion of $3$ always select $(p,q)$-veto core candidates, where $p$ is uniform over all voters and each $q(c)$ is proportional to the plurality score of candidate $c$, i.e., proportional to the number of voters that rank $c$ first. 
However, our understanding of distortion in the veto core is restricted to this specific pair of weight vectors and an important open problem arises, as highlighted by \citet{KK23}: ``The study of the generalized veto core and its connection to metric distortion is very intriguing. In particular, for which initial candidate weights $q$ do candidates in the core guarantee constant (or otherwise small) distortion?'' Our first result makes significant progress toward resolving this open problem by establishing upper bounds on the distortion of candidates in the 
$(p,q)$-veto core for arbitrary choices of both $p$ and $q$. We then exhibit the power of these bounds by revisiting the metric distortion problem through the learning-augmented framework and leveraging our bounds to design optimal learning-augmented algorithms.

\paragraph{{\bf Learning-augmented framework.}}
Aiming to overcome the limitations of worst-case analysis, a surge of recent work has focused on a new analysis framework for achieving refined bounds using the guidance of (machine-learned) predictions. In this framework, often termed \emph{algorithms with predictions} or \emph{learning-augmented algorithms}, the algorithm is enhanced with a prediction that it can use as a guide toward improved performance. This prediction can take different forms (depending on the information that the designer may benefit from in the setting at hand) and can be generated using historical data or other relevant information available to the designer.
However, the quality of this prediction is unknown and cannot be trusted.
This framework goes beyond worst-case analysis through bicriteria optimization: the algorithm is simultaneously evaluated based on its performance when the prediction is accurate (known as its \emph{consistency}) as well as its performance when the prediction can be arbitrarily inaccurate (known as its \emph{robustness}). 

One extreme choice is to follow the prediction blindly, which is rewarding when it is accurate (leading to good consistency), but can yield very bad outcomes when it is inaccurate (leading to poor robustness). The other extreme is to disregard the prediction, which provides weak consistency guarantees. In general, each learning-augmented algorithm provides a trade-off between robustness and consistency, and the goal is to identify the Pareto frontier between these two measures.

The learning-augmented framework has been applied quite broadly, {\sl e.g.}, to analyze the competitive ratio of online algorithms, the running time of algorithms and data structures, or the performance of mechanisms in multiagent systems
(see~\cite{alps} for a frequently updated list of papers in this rapidly growing literature). In all of these applications, this framework has been used to mitigate information limitations that pose obstacles for the designer.

Since the main challenge in metric distortion is the information gap that the designer faces, this makes it a prime application domain for this framework.
Specifically, apart from the rankings of the voters, it is reasonable to assume that the designer may also have access to historical observations regarding the voters' choices in other matters that may correlate with their preferences in the matter at hand, thus providing hints regarding their preferred outcome in the metric space.

\subsection{Our Results}
Our first main result provides an upper bound for the distortion of any algorithm choosing candidates from the $(p,q)$-veto core, for arbitrary weight vectors $p\in \Delta(\V)$ and $q \in \Delta(\C)$.

\vspace{0.1in}
\noindent {\bf Theorem:}
Any algorithm choosing a candidate $c$ from the $(p,q)$-veto core has distortion at most 
$$
1+\frac{2 \max_vp(v)}{\min_{c} \frac{q(c)}{\plu(c)}},
$$
where $\plu(c)$ is the plurality score of $c$ (the number of voters that rank $c$ at the top of their list).

Note that all the deterministic algorithms known to achieve the optimal distortion of 3 choose candidates from the $(\unip,q^\plu)$-veto core, where $\unip(v)=\frac{1}{n}$ for all $v$ and $q^\plu(c)=\frac{\plu(c)}{n}$ for all $c$. For these vectors, our bound recovers the optimal distortion of 3. In general, our upper bound makes significant progress towards answering the question posed by~\citet{KK23} by providing a sufficient condition for achieving constant distortion, i.e., that the ratio of $\max_vp(v)$ over $\min_c \frac{q(c)}{\plu(c)}$ is constant. Another implication of our upper bound is that if we let both $p$ and $q$ be uniform, in which case the $(p,q)$-veto core reduces to the \emph{proportional veto core} introduced by~\citet{Moulin81}, the resulting distortion is at most $1+2\frac{m}{n}\max_{c\in \C}\plu(c)$. 
Intuitively, $\max_vp(v)$ captures the ``distance'' of $p$ from $\unip$ (for which $\max_v \unip(v)=\frac{1}{n}$) and $\min_c \frac{q(c)}{\plu(c)}$ captures the ``distance'' of $q$ from $q^{\plu}$ (for which $\min_c q^{\plu}(c)/\plu(c)=\frac{1}{n}$). If we use $\lambda = n \cdot \max_v p(v)$ and $\mu = n \cdot \min_c q(c)/\plu(c)$ as two parameters that quantify these ``distances,'' our upper bound becomes $1+\frac{2\lambda}{\mu}$, and we complement it by showing that it is tight with respect to $\mu$ and $\lambda$. 

Our second main result, leverages our distortion upper bounds for the $(p,q)$-veto core, to provide the first analysis of algorithms for the metric distortion problem using the learning-augmented framework. We first consider learning-augmented algorithms that are enhanced with the strongest possible type of prediction, i.e., a prediction regarding the exact cardinal distances between each voter-candidate pair, and we exhibit an inevitable trade-off between the optimal robustness and consistency pairs that any such algorithm can achieve
(see Theorem \ref{thm:mainlowerbound}). We then propose a family of algorithms augmented only with a prediction $\p\in \C$ regarding who the optimal candidate is, and we prove that despite the weaker prediction that they are provided with, these algorithms are able to match the optimal robustness-consistency trade-off that even the stronger class of algorithms can achieve (see Theorem \ref{thm:optimal_algorithm}). These algorithms, termed $\SimVeto$, are parameterized by $\delta\in [0,1)$ and they always return a candidate from the $(p,q)$-veto core, where $p$ is uniform and $q$ is guided by the prediction, assigning a higher amount of weight to $\hat{c}$; the exact amount also depends on the value of $\delta$ chosen by the designer.

\vspace{0.1in}
\noindent {\bf Theorem:}    For any $\delta \in \left[0,1\right)$, $\SimVeto$ achieves $\frac{3-\delta}{1+\delta}$-consistency and $\frac{3+\delta}{1-\delta}$-robustness.
Moreover, this  is the optimal trade-off.
Namely, no deterministic algorithm that is $\frac{3-\delta}{1+\delta}$-consistent can be strictly better than $\frac{3+\delta}{1-\delta}$-robust, even for the line metric and just two candidates. 
\vspace{0.1in}

\begin{figure}[h]
    \centering
    \includegraphics[width = 0.48 \textwidth]{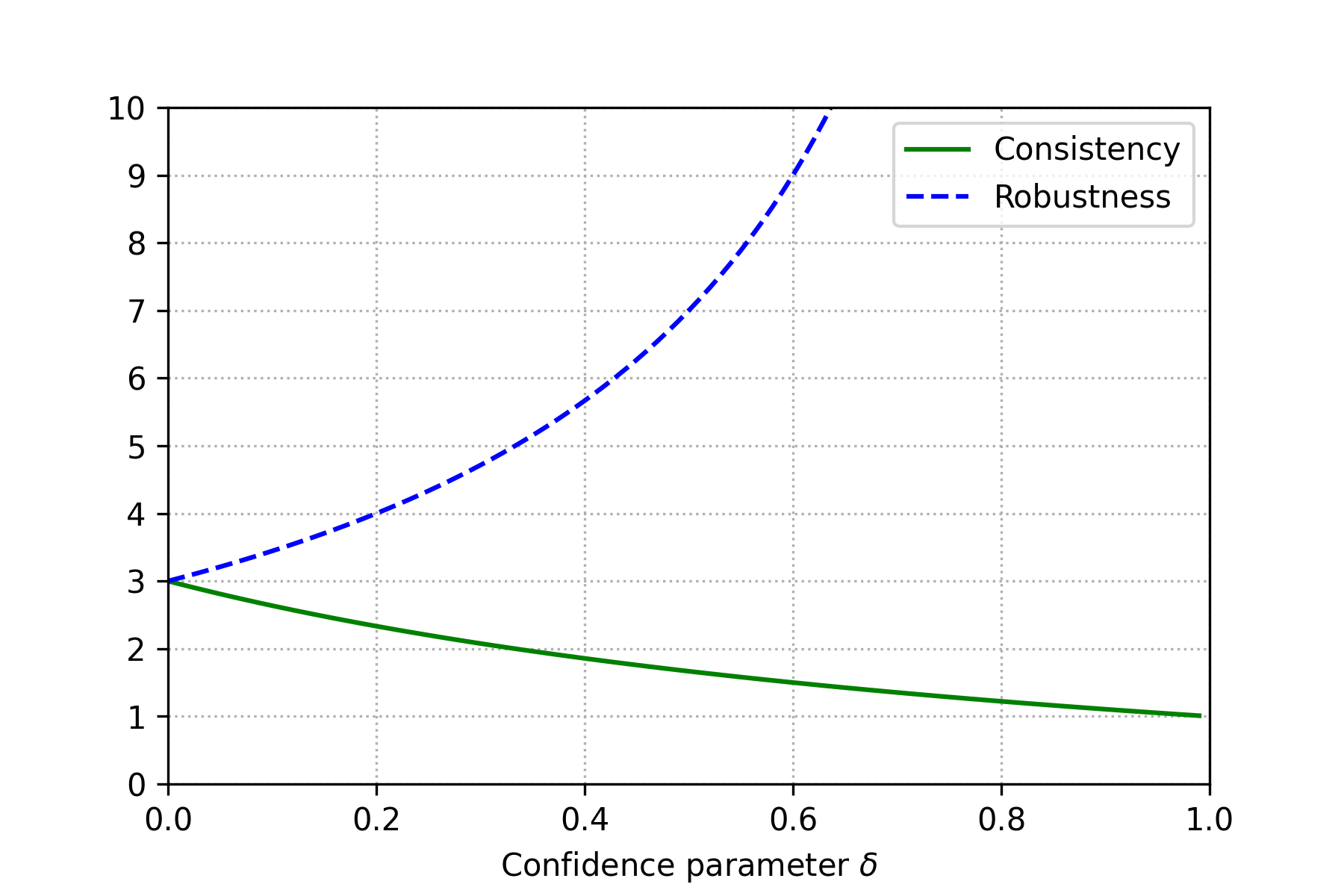}
    \includegraphics[width = 0.48 \textwidth]{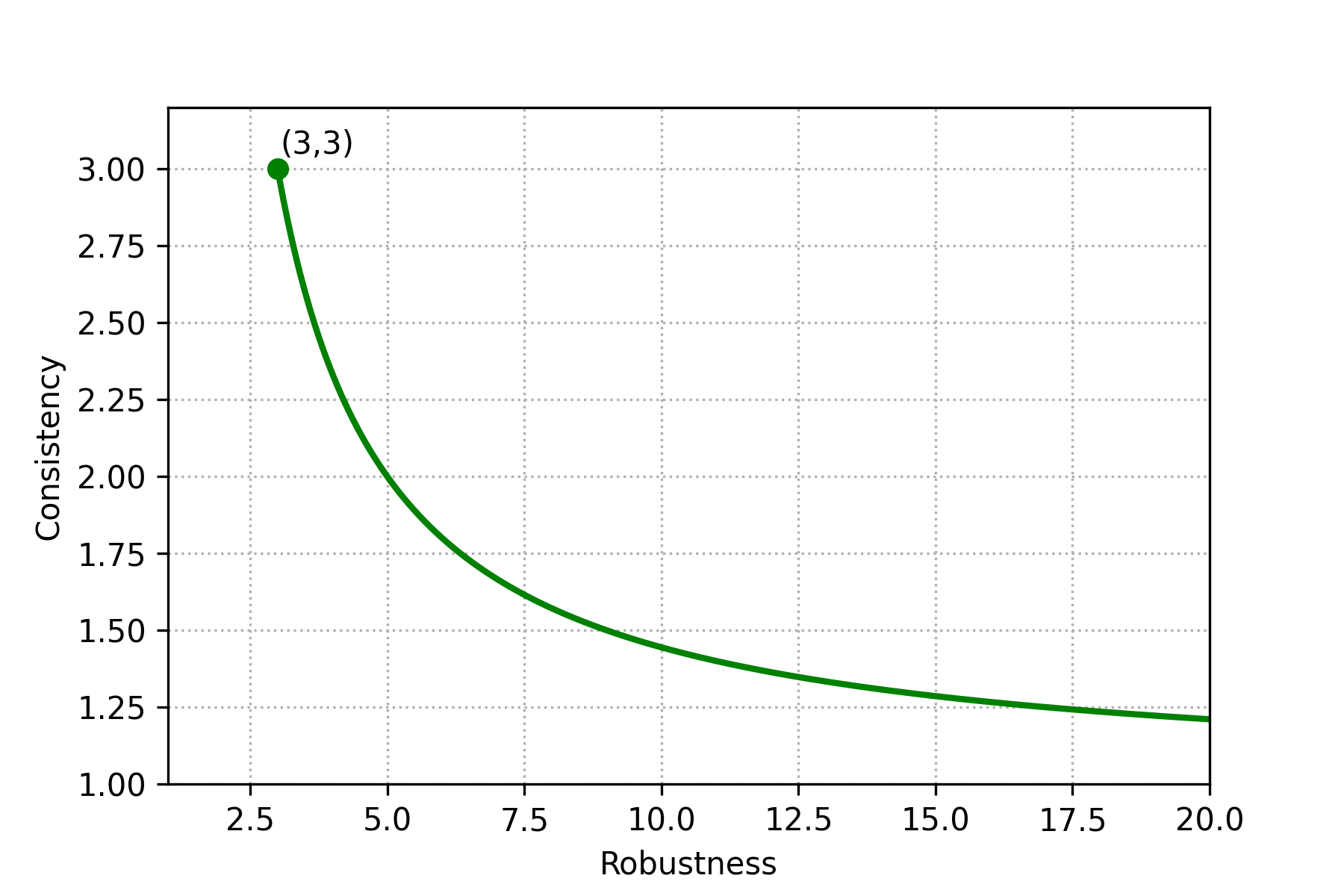}
    \caption{On the left, a plot showing the consistency and robustness bounds achieved by $\SimVeto$ as a function of the confidence parameter $\delta$. On the right, a plot of the Pareto-frontier capturing the best possible trade-off between robustness and consistency, all of which can be achieved by $\SimVeto$ for the appropriate choice of $\delta$.}
    \label{fig:optimal_trade-off}
\end{figure}

Figure~\ref{fig:optimal_trade-off} exhibits this optimal trade-off for values of $\delta\in [0,1)$. 
Specifically, if the designer is confident regarding the quality of the prediction and willing to relax the robustness guarantee to $\beta> 3$, then choosing the corresponding value of $\delta$ yields a consistency guarantee of $(\beta+3)/(\beta-1)$. For example, the designer can achieve a consistency of $2$ with a guaranteed robustness of $5$.

\paragraph{{\bf A more refined analysis.}}
Consistency and robustness capture two extremes regarding the quality of the prediction, namely, full accuracy and complete arbitrariness, respectively. A more refined analysis provides bounds on the distortion parameterized by the quality of the prediction. In Section~\ref{sec:error}, we provide such refined worst-case bounds on the distortion of $\SimVeto$ algorithms.
Specifically, we define the prediction error as $\error\vcentcolon =\frac{n \cdot d(\p, \o)}{\SC(\o)}$, {\sl i.e.}, the distance between the predicted and the optimal candidates, normalized\footnote{As we discuss in Section~\ref{sec:error}, normalizing the error using the average distance from the optimal candidate is necessary to accurately quantify the quality of the prediction.} by the optimal average distance, $\SC(\o)/n$. 
We then prove that the distortion of $\SimVeto$ with any given $\delta$ is at most the bound shown below, which recovers the consistency guarantee of $(3-\delta)/(1+\delta)$ for $\error=0$, then increases linearly in $\error$ up until it reaches the robustness bound of $(3+\delta)/(1-\delta)$. 
 $$\min \left\{\frac{3-\delta+2\delta \error}{1+\delta},~ \frac{3+\delta}{1-\delta}\right\}.$$

\paragraph{{\bf Metric spaces with decisive voters.}}
We then refine the distortion bounds further, using the $\alpha$-decisiveness framework of \citet{AP17}. For $\alpha \in [0,1]$, a voter is $\alpha$-decisive if 
her distance from her favorite candidate is at most $\alpha$ times her distance from any other candidate. 
An instance is $\alpha$-decisive if every voter is $\alpha$-decisive. We obtain a distortion upper bound of 
$$\frac{2 + \alpha -\alpha \delta}{1+\delta} + \min \left\{\frac{2\delta \error}{1+\delta},~ \frac{8\delta}{(1+\delta)(1-\delta)}\right\}$$
for $\alpha$-decisive instances, which recovers our previous bounds when $\alpha=1$.
For $\alpha = 0$, the distance from each voter to her favorite candidate is $0$, capturing the well-studied \emph{peer selection} problem, where the set of voters and the set of candidates coincide and each voter ranks herself first.

\subsection{Related Work}\label{sec:related_work}
\paragraph{Deterministic metric distortion.} \citet{anshelevich2018metric} initiated the study of the metric distortion problem and showed that well-known algorithms such as ``plurality'', ``Borda count'', ``$k$-approval'', and ``veto'' admit a distortion that increases linearly with the number of voters or candidates. They also showed that the distortion of the transferable vote algorithm grows logarithmically in the number of candidates, while Copeland's rule achieves a
distortion of $5$.
They also proved a lower bound of 3 for any deterministic algorithm and conjectured that this bound is tight.

The first deterministic algorithm to guarantee a distortion better than $5$ was proposed by \citet{munagala2019improved}, who achieved a distortion of $2+\sqrt{5} \approx 4.236$ using a novel algorithm.
They also provided a sufficient condition for achieving a distortion of $3$.
Independently, \citet{kempe2020duality} derived the same bound of $2+\sqrt{5}$ using a linear programming duality framework, and provided several alternative formulations of the sufficient condition outlined in \cite{munagala2019improved}.
\citet{GHS20} then introduced the Plurality Matching rule that achieves the optimal distortion of $3$.

Even after the tight distortion bound of $3$ was achieved, subsequent work provided a better understanding of the problem. 
\citet{KK22} achieved the optimal distortion with the Plurality Veto algorithm which is much simpler relative to Plurality Matching,
and very recently  \citet{KK23} proposed a refinement of Plurality Veto called Simultaneous Veto that resolves the arbitrary tie-breaking issues in \citep{GHS20} and \citep{KK22} while maintaining optimal distortion.

\citet{GLS23} studied the extent to which metric distortion bounds can be combined with distortion bounds for the utilitarian setting --- where the voters' preferences are arbitrary normalized valuations.
\citet{abramowitz2019awareness} studied a variant where more information besides the usual preference rankings is available,
and \citet{kempe2020communication} studied the trade-offs between the achievable distortion and the communication complexity of deterministic social choice algorithms.

\paragraph{Metric distortion with randomized algorithms.} The tight bound of $3$ does not hold when algorithms are allowed to be randomized.
\citet{AP17} proved that the Random Dictatorship algorithm
achieves distortion better than $3$, but that tends to 3 as the number of voters grows.
Subsequent work by \citet{fain2019random} and \citet{kempe2020communication} identify other algorithms that achieve distortion bounds that tend to 3 (but this time as the number of candidates grows). 
A lower bound of 2 was also established by \cite{AP17}, which was improved to 2.1126 and 2.06316, independently by \citet{CR22}, and \citet{ps21} respectively.
 \citet{CWRW24} also achieved an upper-bound of 2.753, thereby providing the first randomized algorithm with distortion bounded away from 3.

Unlike the deterministic metric distortion problem, the randomized variant remains unresolved even from the traditional worst-case approach. Closing the remaining gap is a major open question in this field and as more advances are achieved in this direction, this provides additional motivation for a refined analysis of that variant through the learning-augmented framework as well.

\paragraph{Learning-augmented framework.}
The learning-augmented framework has been widely accepted in recent years as a valuable paradigm for the design and analysis of algorithms, 
aiming to circumvent overly-pessimistic worst-case bounds. 
In the last five years alone,
more than 150 papers have revisited
classic algorithmic problems using this framework, with principal examples being online paging \citep{lykouris2018competitive},  scheduling \citep{KPZ18}, secretary problems \citep{dutting2021secretaries,AGKK20}, optimization problems with covering \citep{BMS20} and knapsack constraints \citep{im2021online}, as well as graph problems \citep{azar2022online}. We refer the reader to \citep{MV21} for a survey of some early work and \citep{alps} for a more up-to-date list of papers in this area. More closely related to this paper, some recent work has used this framework on social choice problems, where the predictions are regarding a group of agents and their preferences, and the goal is to optimize some social cost or social welfare function. This includes the study of resource allocation problems where agents' preferences are revealed in an online fashion \citep{banerjee2020online,BGHJMS23} and settings where the agents are strategic and their preferences may be private \citep{ABGOT22,XL22,GKST22,BGT23,IB22,BPS23, LWZ24,BGTZ24,CK24}. Our work adds to this literature by focusing on settings where access to the agents' preferences is limited to ordinal information. 

\paragraph{Metric spaces with $\alpha$-decisive voters.}
The $\alpha$-decisiveness framework was introduced by \citet{AP17} to provide a more refined analysis of distortion bounds, as a function of how strong the agents' preference is for their top choice, and it has been used in several distortion problems
\cite{AFP22,GAX17,FG22}.
The special case of $\alpha = 0$ 
corresponds to 
the \emph{peer selection} setting, where the set of voters and candidates coincide. 
This setting has received attention in settings beyond distortion as well
\cite{alon2011sum, indyk1999sublinear, chang2012some, chang2017lower, chang2017metric}.

\section{Model and Preliminaries}
\label{sec:preliminaries}
Let $(\X,\d)$ be a metric space where $\d:\X\times\X \rightarrow \R_{\geq 0}$ is the distance function that satisfies the following conditions: 1) positive definiteness: $\forall a,b \in \X$, $d(a,b) \geq 0$ and $d(a,b) = 0$ if and only if $a = b$; 2) symmetry: $\forall a,b \in \X$, $d(a,b) = d(b,a)$; and 3) triangle inequality: $\forall a,b,c \in \X$, $d(a,b) + d(b,c) \geq d(a,c)$. 
Let
$\V,\C$
be finite sets of voters and candidates, respectively, that are located in the metric space. We denote by $n$ the number of voters ($n = \left|\V\right|$) and by $m$ the number of candidates ($m = \left|\C\right|$).
For simplicity of notation, we extend $d$ to also operate on the voters and the candidates.
That is, we use $d(a,b)$, for $a,b \in \V \cup \C$, to denote the distance between the points in $\X$ where $a$ and $b$ are located. 
Given a metric $d$, the \emph{social cost} of a candidate $c\in \C$ is $\SC(c,d) = \sum_{v \in \V} d(v,c)$; we write $\SC(c)$ when the metric distance $d$ is clear from the context.

\paragraph{Preference profile.}
We refer to triplets $(\V,\C,\d)$ as \emph{instances} and the distance between a voter and a candidate 
quantifies how much the voter prefers that candidate --- the closer the better. 
We say that a voter $v \in \V$ prefers $c\in \C$ over $c' \in \C$, and we write $c \pref_v c'$ if and only if $\d(v,c) \leq \d(v,c')$.  Note that this is a weak preference relation.
A given instance $(\V,\C,\d)$ induces preference rankings over candidates for each voter.  That is, for each $v \in \V$ we have a preference ranking (bijection) $\sigma_v:\C \rightarrow \{1,\ldots,|\C|\}$ such that $\sigma_v(c_1) < \sigma_v(c_2)$ implies $c_1 \pref_v c_2$, 
and we 
write
in this case that 
$\d$ is \emph{aligned} 
with $\sigma_v$, denoted 
$\d \: \triangleright \sigma_v$. 
Note that these rankings are not determined uniquely when there are at least two candidates whose distance from $v$ is the same.

A \emph{preference profile} $\sigma := (\sigma_v)_{v \in \V}$ is a tuple of preference rankings for each voter and we say that 
$d$ is \emph{aligned}
with the preference profile $\sigma$, denoted as $d \: \triangleright\: \sigma$, if $d \: \triangleright\: \sigma_v$ for each $v \in \V$.
Given $\sigma$, we define $\topc(v)$ to be the candidate ranked highest by $v$, and we define $\plu(c)$ to be the number of voters that rank $c$ highest.

\paragraph{Metric distortion.}
In the metric distortion problem, an algorithm $\alg$ receives as input a preference profile $\sigma$ which is induced by some $(\V,\C,\d)$.
Importantly, $\alg$ does not have access to the underlying distance function.
The goal is to output a candidate that minimizes the social cost.
We denote by $\o(\d) = \arg\min_{c \in \C} \SC(c,\d)$ the candidate that minimizes the social cost, {\sl i.e.}, the optimal candidate,
breaking ties arbitrarily.
The \emph{distortion} of $\alg$ is the worst-case multiplicative approximation it achieves to the optimum, {\sl i.e.},
\[
\dist(\alg) = \sup_{\sigma}\ \sup_{d\: :\: d\: \triangleright\: \sigma}\ \frac{\SC(\alg(\sigma),d)}{\SC(\o(d),d)}.
\]
In addition to using $\o$ to represent the optimal candidate, we often use $a$ to represent a special candidate with some properties and $c$ to refer to any general candidate.

\paragraph{Generalized veto core}
Given a preference profile $\sigma$, a $(p,q)$-domination graph for some candidate $a$ is a vertex-weighted bipartite graph with voters on one side and candidates on the other side. It captures how the candidate is evaluated compared with the other candidates. More specifically, an edge $(v,c)$ between some voter $v$ and candidate $c$ exists if voter $v$ prefers candidate $a$ over candidate $c$. The more edges in the $(p,q)$-domination graph of some candidate, the more ``popular'' she is among the voters. We use $\Delta(\V)$ and $\Delta(\C)$ to denote the probability distribution over $\V$ and $\C$, respectively.
\begin{definition}[$(p,q)$-Domination Graph \cite{GHS20}]\label{def:pqgraph} Given a preference profile $\sigma$ and (normalized) weight vectors $p \in \Delta(\V)$ and $q \in \Delta(\C)$, the \emph{$(p,q)$-domination graph} of candidate $a$ is the vertex-weighted bipartite graph $G^\sigma_{p,q}(a) = (V,C,E_a,p,q)$, where $(v,c) \in E_a$ if and only if $a \pref_v c$. Vertex $v \in V$ has weight $p(v)$ and vertex $c \in C$ has weight $q(c)$. 
\end{definition}

\begin{definition}[Fractional perfect Matching]\label{def:fractionalperfectmatching}
We say that the $(p,q)$-domination graph of candidate $a$, i.e., $G^\sigma_{p,q}(a)$, admits a \emph{fractional perfect matching} if there exists a weight function $w: E_a \rightarrow \R_{\geq 0}$ such that the total weight of edges incident on each vertex equals the weight of the vertex, i.e., for each $v \in V$ we have $\sum_{c \in C: (v,c) \in E_a} w((v,c)) = p(v)$, and for each $c \in C$ we have $\sum_{v \in V: (v,c) \in E_a}w((v,c)) = q(c)$. 
\end{definition}
\citet{KK23} generalized the notion of proportional Veto Core introduced by \citet{Moulin81} by defining the $(p,q)$-veto core and they showed that it is equivalent to the set of candidates whose $(p,q)$-domination graph admits a fractional perfect matching. 

\begin{definition}[$(p,q)$-Veto Core \cite{KK23}]\label{def:pqvetocore} Given a preference profile $\sigma$ and weight vectors $p \in \Delta(\V)$ and $q \in \Delta(\C)$, the $(p,q)$-Veto Core is the set of candidates $a$ such that $G^\sigma_{p,q}(a)$ admits a fractional perfect matching.
\end{definition}
\section{Distortion bounds for $(p,q)$-Algorithms}\label{sec:pqsection}
In this section, we prove distortion bounds for algorithms that take as input a preference profile $\sigma$ and return candidates from the $(p,q)$-veto core of $\sigma$. Specifically, in Section~\ref{sec:unifromp_generalq} we establish bounds for arbitrary $q \in \Delta(\C)$ paired with a uniform $p$ vector. We then extend this bound in Section~\ref{sec:generalp_generalq} to arbitrary $p \in \Delta(\V)$.

We henceforth refer to the class of algorithms that we will be analyzing as $(p,q)$-algorithms.

\begin{definition}[$(p,q)$-algorithms]\label{def:pqalgorithm}
Given arbitrary weight vectors $p \in \Delta(\V)$ and $q \in \Delta(\C)$, a $(p,q)$-algorithm is one that,
given a preference profile $\sigma$ as input, returns a non-empty subset of candidates from the $(p,q)$-veto core of $\sigma$.
\end{definition}

The initial motivation to study such algorithms is due to the work of \citet{munagala2019improved}, who focused on the special case of weight vectors $(\unip, q^\plu)$, where $\unip(v) = 1/n$ for all $v \in \V$ and $q^\plu(c) = \plu(c)/n$ for all $c \in \C$, and showed that if the $(\unip, q^\plu)$-veto core is non-empty, then any $(\unip, q^\plu)$-algorithm achieves an optimal distortion of $3$.
\begin{lemma}[\cite{munagala2019improved,GHS20}] If the $(\unip, q^\plu)$-veto core is non-empty, then the distortion of any $(\unip, q^\plu)$-algorithm is at most 3.
\end{lemma}
Building upon this lemma, \citet{GHS20} achieved the optimal distortion bound by proving that the $(p, q)$-veto core of any preference profile is non-empty for any choice of $p$ and $q$.
\begin{lemma}[Ranking-Matching Lemma \cite{GHS20}]\label{lem:ranking_matching_lemma}
For any preference profile
$\sigma$ and weight vectors $p \in \Delta(\V)$ and $q \in \Delta(\C)$, there exists a candidate $a$ whose $(p,q)$-domination graph $G^\sigma_{p,q}(a)$ admits a fractional perfect matching. Namely, the $(p,q)$-veto core is non-empty.
\end{lemma}

The $(p,q)$-algorithm proposed by \citet{GHS20} (\textsc{PluralityMatching}) identifies an arbitrary candidate in the $(p,q)$-veto core by just iterating over all candidates and checking whether their $(p,q)$-domination graph has a fractional perfect matching. Subsequent work by \citet{KK22,KK23} proposed
more elegant $(p,q)$-algorithms that return some subset of candidates in the $(p,q)$-veto core (e.g., \textsc{SerialVeto} and \textsc{SimultaneousVeto}). However, our understanding of the performance of these algorithms with respect to distortion remains limited to the special case of the $(\unip, q^\plu)$ weight vectors. Our first result makes significant progress on this matter by establishing a distortion upper bound for $(p,q)$-algorithms with
arbitrary $(p,q)$ weights.

\subsection{Distortion Bounds for $\unip$ and General $q$}\label{sec:unifromp_generalq}
In this section, we provide distortion bounds for $(\unip,q)$-algorithms with arbitrary $q$. Given any preference profile $\sigma$ (inducing $\plu(c)$ for each candidate $c\in \C$) and any weight vector $q \in \Delta(\C)$, our analysis utilizes the following parameter: 
 \begin{align}\label{eq:mdef}
        \m(\sigma, q) = \min_{c \in \C} \frac{q(c) \cdot n}{\plu(c)}.
    \end{align}
We simply write $\mu$ when $\sigma$ and $q$ are clear from the context, and we observe (by an averaging argument) that $\mu \leq 1$. 
We note that any candidate with $\plu(c)=0$ can be treated as if $q(c)/\plu(c)=1$ when computing this parameter value.
\begin{theorem}\label{thm:unip_genq}
Given any $q \in \Delta(\C)$, the distortion of any $(\unip, q)$-algorithm is at most $1+\frac{2}{\mu}$.
\end{theorem}
At a high level, for a given preference profile $\sigma$, $\m$ quantifies the distance between the given vector $q$ and the vector $q^\plu$ induced by the instance. More formally, $\m$ represents the fraction of the weight in $q$ that aligns with the plurality score. If this part of weight were the only component present 
then the distortion for any candidate within the corresponding veto core would be at most $3$. 
Consequently, we decouple the weight into two components and analyze them separately. For the ``well-behaved'' $\m$ portion of the weight, we utilize the analysis established in prior work. Addressing the ``ill-behaved'' weight, which comprises the remaining $1-\m$ fraction, necessitates a deeper understanding of the $(p,q)$-domination graph and the implication of fractional perfect matching on the quality of the corresponding candidates.

To prove Theorem~\ref{thm:unip_genq}, {\sl i.e.}, the upper bound on $\SC(a,d)/\SC(\o,d)$ for any agent $a \in (\unip,q)$-Veto Core in any metric $d$, the key technical contribution of this section is 
establishing an upper bound on the distance between the optimal candidate, $\o$, and the corresponding candidate, $a$, as a function of the optimal social cost (see Lemma~\ref{lem:opt_bound_matchinganyq}). To prove this upper bound, we extend the domination graph of $a$, $G^\sigma_{\unip,q}(a)$, to a Voter Domination Graph (\vdg~graph, see below) for $a$ that captures the conflicts among voter preferences. 
Leveraging the structure of this graph, we derive a lower bound on the distance between pairs of voters from the optimal candidate, $\o$, as stated in Lemma~\ref{lem:keylemmav2anyq}.

To this end, we first define the \emph{Voter Domination Graph (VDG)} of some candidate $a$. 
\begin{definition}\label{def:domination_graph}
Let $\sigma$ be a preference profile, $a \in \C$ be a candidate, and $p \in \Delta(\V)$, $q \in \Delta(\C)$ be weight vectors such that $G^\sigma_{p,q}(a)$ admits a fractional perfect matching. We define the weight vector $p' \in\Delta(\V \cup \{b\})$, where $p'(v) = \frac{\mu}{n}$ and $p'(b) = 1-\mu$, and the \emph{voter's domination graph} (VDG) of $a$ as a bipartite graph $\vdg^\sigma_{p,p'}(a) = (V, V \cup \{b\}, E'_a)$, where $(v,v') \in E'_a$ if and only if $a \pref_v \topc(v')$, and $(v, b) \in E'_a$ if $(v,c) \in E_a$ and $q(c) > \mu \frac{\plu(c)}{n}$. We simply write $\vdg(a)$ when the parameters are clear from the context.
\end{definition}
\begin{figure}
	\begin{subfigure}[b]{0.29\textwidth}
		\centering
		\[\begin{array}{ll}
			  \sigma_1 :& a \succ b \\
			  \sigma_2 :& b \succ a \\
			  \sigma_3 :& b \succ a 
		\end{array}\]
		\caption{Preference Profile $\sigma$}
		\label{fig:rankings}
	\end{subfigure}
	\begin{subfigure}[b]{0.29\textwidth}
		\centering
		\begin{tikzpicture}[scale=.75]
			\tikzstyle{main_node} = [circle,fill=white,draw,minimum size=1.2em,inner sep=0pt]

			\node[main_node] (1) at (0, 0) {$1$};
			\node[main_node] (2) at (0, -1) {$2$};
			\node[main_node] (3) at (0, -2) {$3$};
			\node[main_node] (a) at (2, -.5) {$a$};
			\node[main_node] (b) at (2, -1.5) {$b$};

			\draw (1) node [left=.5em] {\tiny $p(1) = \frac{1}{3}$};
			\draw (2) node [left=.5em] {\tiny $p(2) = \frac{1}{3}$};
			\draw (3) node [left=.5em] {\tiny $p(3) = \frac{1}{3}$};

			\draw (a) node [right=.5em] {\tiny $q(a) = \frac{2}{3}$};
			\draw (b) node [right=.5em] {\tiny $q(b) = \frac{1}{3}$};



   		\draw (1)--(b);
            \draw (1)--(a);
			\draw (2)--(a);
			\draw (3)--(a);
			\draw (1)--(b) [draw=red, line width=.75mm];
			\draw (2)--(a) [draw=red, line width=.75mm];
			\draw (3)--(a) [draw=red, line width=.75mm];
		\end{tikzpicture}
		\caption{$G^\sigma_{\unip, q}(a)$}
		\label{fig:a-domination}
	\end{subfigure}
	\begin{subfigure}[b]{0.29\textwidth}
		\centering
		\begin{tikzpicture}[scale=.75]
			\tikzstyle{main_node} = [circle,fill=white,draw,minimum size=1.2em,inner sep=0pt]

			\node[] () at (-.8, 0) {};

			\node[main_node] (1l) at (0, 0) {$1$};
			\node[main_node] (2l) at (0, -1) {$2$};
			\node[main_node] (3l) at (0, -2) {$3$};

            \draw (1l) node [left=.5em] {\tiny $p(1) = \frac{1}{3}$};
			\draw (2l) node [left=.5em] {\tiny $p(2) = \frac{1}{3}$};
			\draw (3l) node [left=.5em] {\tiny $p(3) = \frac{1}{3}$};

			\node[main_node] (br) at (2, 1) {$b$};
			\node[main_node] (1r) at (2, 0) {$1$};

			\node[main_node] (2r) at (2, -1.5) {$2$};
            \node[main_node] (3r) at (2, -2.5) {$3$};

            \draw (br) node [right=.9em] {\tiny $p'(b) = \frac{1}{2}$};
            \draw (1r) node [right=.9em] {\tiny $p'(1) = \frac{1}{6}$};
			\draw (2r) node [right=.9em] {\tiny $p'(2) = \frac{1}{6}$};
			\draw (3r) node [right=.9em] {\tiny $p'(3) = \frac{1}{6}$};


			\draw[densely dotted] (1.5, -0.5)--(2.5,-0.5)--(2.5,1.5)--(1.5,1.5)--cycle;

			\draw[densely dotted] (1.5, -3)--(2.5,-3)--(2.5,-1)--(1.5,-1)--cycle;

			 \draw (1l)--(br);
			 \draw (1l)--(1r);
			 \draw (1l)--(2r);
			 \draw (1l)--(3r);
			 \draw (2l)--(1r);
			 \draw (3l)--(1r);
			 \draw (1l)--(2r) [draw=red, line width=.5mm];
              \draw (1l)--(3r) [draw=red, line width=.5mm];
			 \draw (2l)--(1r) [draw=red, line width=.25mm];
              \draw (3l)--(1r) [draw=red, line width=.25mm];
              \draw[dotted] (2l)--(br) [draw=red, line width=.75mm];
              \draw[dotted] (3l)--(br) [draw=red, line width=.75mm];
		\end{tikzpicture}
		\caption{$\vdg(a)$, $\mu = 0.5$}
		\label{fig:a-integral}
	\end{subfigure}
	\caption{
 An example of a domination graph and a \vdg~graph for an instance with preference profile $\sigma$. The red edges represent a matching. The value of $\mu(q,\sigma)$ is $\frac{q(b) \cdot 3}{\plu(b)} = \frac{1}{2}$. In the graph $G^\sigma_{\unip, q}(a)$, the weight of each edge in the matching is $1/3$. In $\vdg(a)$, $w'((2,1)) = w'((3,1)) = 1/12$, $w'((1,2)) = w'((1,3)) = 1/6$, and $w'((3,b)) = w'((2,b)) = 1/4$. The dotted edges in $\vdg(a)$ represent the ``ill-behaved'' portion of the weight.}
	\label{fig:matching-example}
\end{figure}
We note that Definition~\ref{def:domination_graph} closely resembles Definition 5 from \cite{GHS20}. The key difference lies in the requirement for the graph described in Definition 5 to be integral, with integrality being essential for the analysis to go through. In contrast, our graph may be fractional.
In addition, we introduce an extra voter $b$ on the RHS to capture the ``ill-behaved'' weights which does not exist in \cite{GHS20}'s setting. See Figure~\ref{fig:matching-example} for an illustration of $\vdg$.
We first observe that for any candidate $a$ whose domination graph $G^\sigma_{p,q}(a)$ admits a fractional perfect matching, $\vdg(a)$ also admits a fractional perfect matching. 
\begin{lemma}\label{obs:matchingtomatchinganyq}
Consider any candidate $a$ such that $G^\sigma_{\unip,q}(a)$ admits a fractional perfect matching for some arbitrary $q$, then $\vdg(a)$ 
also admits a fractional perfect matching, i.e., there exists a weight function $w': E'_a \rightarrow \R_{\geq 0}$ such that for each voter $v \in \V$ we have $\sum_{v' \in \V \cup \{b\}: (v,v') \in E'_a}w'((v,v')) = p(v)$ and for each $v' \in \V \cup\{b\}$ we have $\sum_{v \in \V: (v,v') \in E'_a}w'((v,v')) = p'(v')$.
\end{lemma}

\begin{proof}
We prove the statement by constructing $w':E_a' \rightarrow \R_{\geq 0}$ from $w$. 
\allowdisplaybreaks
\begin{align}
    w'((v,v')) &= \frac{\m}{q(\topc(v'))n} \cdot w((v,
    \topc(v')
    ))\label{eq:voterweight}\\
    w'((v,b)) &= \sum_{c\in \C: (v,c) \in E_a} 
    \left(
    1 - \frac{\m \cdot \plu(c)}{q(c)n} 
    \right)
    \cdot w((v,c)) \label{eq:boostweight}
\end{align}
First consider any voter $v$ on the LHS, we have:
\begin{align*}
    \sum_{v':(v,v') \in E'_a} w'((v,v')) & = \left(\sum_{v' \in \V: (v,v') \in E'_a} w'((v,v'))\right) + w'((v,b))\\
    & = \left(\sum_{c:(v,c) \in E_a }\sum_{v':\topc(v') = c} w'((v,v'))\right) + 
    w'((v,b))\\
    & =\sum_{c:(v,c) \in E_a} \frac{\m\cdot \plu(c)}{q(c)n}\cdot w((v,c)) + \sum_{c:(v,c)\in E_a } \left(1 - \frac{\m \cdot \plu(c)}{q(c)n}\right)\cdot w((v,c))\\
    & = \sum_{c:(v,c)\in E_a} \left(1-\frac{\m \cdot \plu(c)}{q(c)n} +\frac{\m \cdot \plu(c)}{q(c)n} \right) \cdot w((v,c))\\
    & = \sum_{c:(v,c) \in E_a} w((v,c)) ~= ~ p(v) \tag{by Def.\ ~\ref{def:fractionalperfectmatching}}
\end{align*}
Now consider any voter $v'$ on the RHS, and let $c = \topc(v')$. We have:
\begin{align*}
    \sum_{v: (v,v') \in E'_a} w'((v,v')) & = \sum_{v:(v,
    c
    ) \in E_a} \frac{\m}{q(
    c
    )n} 
    w((v,c))
    \\
    & =\frac{\m}{q(
    c
    )n} \sum_{v:(v,
    c
    ) \in E_a} 
    w((v,c))
    \\
    & = \frac{\m}{q(
    c
    )n} \cdot q(
    c
    ) \tag{by Def.\ ~\ref{def:fractionalperfectmatching}}\\
    & = \frac{\mu}{n}~=~
    p'(v')
\end{align*}
Finally, we consider the added voter $b$ on the RHS. We have:
\allowdisplaybreaks
\begin{align*}
    \sum_{v: (v,b) \in E'_a} w'((v,b)) & = 
    \sum_{v: (v,b) \in E'_a}
    \sum_{
    c: (v,c) \in E_a
    } 
    \left(1 - \frac{\m \cdot \plu(c)}{q(c)n}\right)\cdot w((v,c))\\
    & = \sum_{c \in \C} \sum_{v:(v,c) \in E_a}\left(1 - \frac{\m \cdot \plu(c)}{q(c)n}\right)\cdot w((v,c))\\
    & = \sum_{c \in \C}\left(1 - \frac{\m \cdot \plu(c)}{q(c)n}\right)\cdot q(c) \tag{by Def.\ ~\ref{def:fractionalperfectmatching}}\\
    &= 1 - 
    \m
    \sum_{c \in \C} \frac{\plu(c)}{n} ~=~ 1-\mu = q'(b) \qedhere
\end{align*}
\end{proof}

The Corollary below follows from Lemma~\ref{obs:matchingtomatchinganyq}.
\begin{corollary}\label{cor:usefulweightsanyq}
$\sum_{(v,v') \in E'_a: v' \in \V} w'((v,v')) = \m$.
\end{corollary}
We first observe the following relations between voters $v,v'$ and candidate $a$, for every pairs of $v,v'$ with $(v,v') \in E_a'$.
\begin{observation}\label{obs:algotooptbound}
Consider any candidate $a$ such that $G^\sigma_{p,q}(a)$ admits a fractional perfect matching for $\unip$ and some arbitrary $q$. If $(v,v') \in E'_a$ in the VDG of candidate $a$, $\vdg(a)$, then for any $d \triangleright \sigma$, $$d(v,a) \leq d(v,\o) +  2d(v',\o).$$
\end{observation}
\begin{proof}
Consider any $v, v' \in \V$ such that $(v,v') \in E'_a$, let $c =\topc(v')$. We have
\begin{align*}
    d(v,a) &\leq d(v,\o) + d(\o,a) \tag{triangle ineq.} \\
            &\leq d(v,\o) + d(\o,c) \tag{by Def.~\ref{def:pqgraph}} \\
            &\leq d(v,\o) + d(\o,v') + d(v',a) \tag{triangle ineq.}\\
            &\leq d(v,\o) + d(v',\o) + d(v',\o) \tag{since $c =\topc(v')$}\\
            & \leq d(v,\o) + 2d(v',\o). \qedhere
\end{align*}
\end{proof}

We now make the crucial observation that if candidate $a$ is in the  $(p,q)$-veto core, then for every edge $(v,v')\in E'_a$ ({\sl i.e.}, for every edge that does not involve $b$ in the VDG of candidate $a$), we can derive a lower bound on the combined contribution of voters $v$ and $v'$ to the social cost of the optimal candidate, $\o$. Intuitively, if $v$ prefers $a$ over the candidate closest to $v'$ (this is the meaning of the assumption that $(v,v') \in E'_a$),
then $v$ must be far enough from that candidate and hence from $v'$ as well.  In turn, if one of $v$ or $v'$ is close to $\o$, then the other voter must be far enough from $\o$.
\begin{lemma}\label{lem:keylemmav2anyq}
Consider any candidate $a$ such that $G^\sigma_{p,q}(a)$ admits a fractional perfect matching for $\unip$ and some arbitrary $q$. If $(v,v') \in E'_a$ in the VDG of candidate $a$, $\vdg(a)$, then for any $d \triangleright \sigma$, $$d(v,\o) + d(v',\o) \geq \frac{d(\o,a)}{2}.$$
\end{lemma}
\begin{proof}
Consider any $v$, by the triangle inequality we have 
\begin{align}\label{eq:triangleinequality}
  d(\o,a) \leq d(a,v) + d(v,\o) ~\Rightarrow~ d(a,v) \geq d(\o,a) - d(v,\o).  
\end{align}
Consider any $v' \in \V$ such that $(v,v') \in E'_a$, let $c = \topc(v')$. We have
\begin{align}\label{eq:candidatesum}
    d(\o, c) ~\geq~ d(c,v) - d(v,\o) ~\geq~ d(a,v) -d(v,\o) ~\geq~ d(\o,a) -2d(v,\o),
\end{align}
where the first inequality holds by the triangle inequality, the second holds since $d(c,v) \geq d(a,v)$ by Definition~\ref{def:domination_graph}, and the last inequality is using~\eqref{eq:triangleinequality}. Now consider voter $v'$, we have
\begin{align*}
 d(\o,v') + d(v',c) &\geq d(\o,c) &\text{(by triangle inequality)} ~~~&\Rightarrow \\
 2d(\o,v') &\geq d(\o,c) &\text{(since $c = \topc(v') \pref_{v'} \o$)}  ~~~&\Rightarrow \\
  2d(\o,v') &\geq d(\o,a) - 2d(v,\o)  &\text{(using Inequality \eqref{eq:candidatesum})}~~~&\Rightarrow\\
  d(\o,v)+d(\o,v') &\geq \frac{d(\o,a)}{2}. \qedhere
\end{align*}
\end{proof}

We now show the key lemma that connects the social cost of the optimal candidate $\o$ with the distance between the candidate $a$ who admits a fractional perfect matching and $\o$. 
\begin{lemma}\label{lem:opt_bound_matchinganyq}
 Consider any candidate $a$ in the $(\unip,q)$-veto core of $\sigma$ for some arbitrary $q$. Then, for any metric $d \triangleright \sigma$,
  \[\SC(\o,d) \geq \left(\frac{\m}{1+\m}n\right) \frac{d(\o,a)}{2}.\]
\end{lemma}
\begin{proof}
Consider the optimal social cost
{\allowdisplaybreaks
\begin{align*} \hspace*{-0.2cm}
    \SC(\o,d) &= \sum_{v \in \V} d(\o,v)\\
            & = \sum_{v \in \V} \frac{1}{p(v)+p'(v)} \cdot (p(v)+p'(v)) \cdot d(\o, v)\\
            & = \frac{n}{1+\m} \sum_{v \in \V} (p(v)+p'(v)) \cdot d(\o, v)\\
             & =\frac{n}{1+\m} \left(\sum_{v \in \V} p(v) \cdot d(\o,v) +\sum_{v' \in \V} p'(v') \cdot d(\o,v')\right) \\
             & = \frac{n}{1+\m}\left(\sum_{v \in \V} \sum_{v' \in \V \cup \{b\}} w((v,v')) \cdot d(\o,v) +\sum_{v' \in \V} \sum_{v \in \V} w((v,v')) \cdot d(\o,v')\right) \tag{by Lem.~\ref{obs:matchingtomatchinganyq}}\\
            & \geq \frac{n}{1+\m}\left(\sum_{v \in \V} \sum_{v' \in \V} w((v,v')) \cdot d(\o,v) +\sum_{v' \in \V} \sum_{v \in \V} w((v,v')) \cdot d(\o,v')\right) \tag{removing $b$}\\
            & \geq \frac{n}{1+\m} \sum_{(v,v') \in E'_a} w'((v,v')) \cdot (d(\o, v) + d(\o, v')) \tag{rearranging the sum}\\
            & \geq \frac{n}{1+\m} \sum_{(v,v') \in E'_a} w'((v,v')) \cdot \frac{d(\o,a)}{2} \tag{by Lem.~\ref{lem:keylemmav2anyq}}\\
            & \geq \frac{n}{1+\m}\cdot \m \cdot  \frac{d(\o,a)}{2}  \tag{by Corollary~\ref{cor:usefulweightsanyq}}\\
            & = \left(\frac{\m}{1+\m}n\right) \frac{d(\o,a)}{2}. \qedhere
\end{align*}}
\end{proof}

We are now ready to prove Theorem~\ref{thm:unip_genq}. 
\begin{proof}[Proof of Theorem~\ref{thm:unip_genq}]
Consider any candidate $a$ in the $(\unip,q)$-veto core of $\sigma$ for $\unip$ and some $q$, i.e., $a$ is any candidate that could be returned by a $(\unip,q)$-algorithm. We show that for any metric $d \triangleright \sigma$,
\[\SC(a,d) \leq \left(1+\frac{2}{\m}\right) \cdot \SC(\o,d),\]
where $\m = \min_{c \in \C} \frac{q(c) \cdot n}{\plu(c)}$.
The social cost of this candidate $a$ can be upper bounded as follows:

 {\allowdisplaybreaks
 \begin{align*}
   \SC(a,d)& = \sum_{v \in \V}d(v,a)\\
        & = \sum_{v \in \V} n \cdot p(v) \cdot d(v,a) \tag{since $p$ is uniform}\\  
        & = n \sum_{v \in \V} \left(\sum_{v' \in \V} w'((v,v')) + w'((v,b))\right) \cdot d(v,a)\\
        &= n \sum_{v \in \V} \sum_{v' \in \V} w'((v,v'))\cdot d(v,a) + n \sum_{v \in \V} w'((v,b)) \cdot d(v,a)\\
        &= n \sum_{v \in \V} \sum_{v' \in \V}  w'((v,v'))\cdot (d(v,\o)+2d(v',\o)) + n \sum_{v \in \V} w'((v,b)) \cdot d(v,a) \tag{by Obs.~\ref{obs:algotooptbound}}\\
        &\leq n \sum_{v \in \V} \sum_{v' \in \V}  w'((v,v'))\cdot (d(v,\o)+2d(v',\o)) + \sum_{v \in \V} w'((v,b)) \cdot (d(v,\o) + d(\o,a)) \\
        & = n \sum_{v\in \V}\sum_{v' \in \V \cup \{b\}}w'((v,v')) \cdot d(v,\o) + n\sum_{v'\in \V}\sum_{v\in \V}w'((v,v')) \cdot 2d(v',\o) + n \sum_{v\in \V} w'((v,b)) \cdot d(\o,a)\\
        & = \sum_{v\in \V}np(v)\cdot d(v,\o) + n\sum_{v'\in \V}np'(v')\cdot 2d(v'.\o) + n(1-\m)\cdot d(\o,a)\tag{by Lem.~\ref{obs:matchingtomatchinganyq}}.\\
        & = \sum_{v\in \V}d(v,\o) + \sum_{v' \in \V} 2\mu\cdot d(v',o) + n(1-\m)\cdot d(\o,a) \tag{by Def.~\ref{def:domination_graph}}\\
        & = (1+2\mu) \SC(\o,d) + n(1-\mu)2\frac{1+\mu}{\m} \SC(\o,d)\\
        & = \left(1+\frac{2}{\m}\right) \cdot \SC(\o,d). \qedhere
 \end{align*}}
 \end{proof}

\subsection{Distortion Bounds for $(p,q)$-algorithms with General $p$ and $q$}\label{sec:generalp_generalq}
We now extend the distortion upper bound we obtained in the previous section for $\unip$ to prove a distortion upper bound for general $p$ (and $q$). Specifically, we reduce any instance with some arbitrary $p$ and $q$ to an instance with $\unip$ and a new $q'$ such that the $(\unip, q')$-veto core is a superset of the original $(p,q)$-veto core. To do so, we show that if --- for some candidate $a$ and arbitrary $p,q$ --- $G^\sigma_{p,q}(a)$ admits a fractional perfect matching, 
then the new $q'$ that we construct has a fractional perfect matching in the domination graph $G^\sigma_{\unip, q'}(a)$, and we carefully bound the change in $\mu$
in the transition from $q$ to $q'$.

The high level idea is to add weights on both sides so as to ``re-balance'' the given $p$ vector, and re-normalize the weight. We introduce a new parameter $\lambda$ to capture the amount of extra weight that (relatively) needs to be added to re-balance $p$. All the proofs missing from this section can be found in Appendix~\ref{app:pqreduction}.
\begin{restatable}{rLem}{lemgenpunipreduction}\label{lem:genp_unip_reduction}
Consider any candidate $a$ in the $(p,q)$-veto core of $\sigma$ for for some arbitrary $p \in \Delta(\V)$ and $q \in \Delta(\C)$, then there exists some $q' \in \Delta(\C)$ such that $G^\sigma_{\unip,q'}(a)$ also admits a fractional perfect matching. In addition, if we let $\lambda = n\max_v p(v)$, we have
\[\mu(\sigma,q') \geq \frac{\mu(\sigma, q)}{\lambda}.\]
\end{restatable}
Combining Theorem~\ref{thm:unip_genq} and Lemma \ref{lem:genp_unip_reduction} the corollary below follows.
\begin{corollary}\label{cor:genp-genq}
Given a preference profile $\sigma$ and any $p \in \Delta(\V)$ and $q \in \Delta(\C)$, the distortion of any $(p,q)$-algorithm is at most $$1+\frac{2\max_v p(v)}{\min_c \frac{q(c)}{\plu(c)}}$$
\end{corollary}

\subsection{Distortion Lower Bounds}\label{sec:pqlowerbound}
We now show that our upper bound is tight with respect to parameters $\m$ and $\lambda$, 
i.e., for any given $\m \in (0,1]$ and $\lambda \in [1,n]$ there exists a metric $d$ and vectors $p \in \Delta(\V)$ and $q \in \Delta(\C)$ that yield the given $\m$ and $\lambda$, such that some candidate $a$ in the $(p,q)$-veto core with distortion at least $1+\frac{2\lambda}{\mu}$. 
\begin{restatable}{rThm}{thmpqlowerbound}\label{thm:tightmlowerbound}
Given any $\m \in (0,1]$ and $\lambda \in [1,n]$,
there exists a preference profile $\sigma$ and weight vectors $q \in \Delta(\C)$ with $\m = \min_{c \in C} \frac{q(c)\cdot n}{plu(c)}$ and $p \in \Delta(\V)$ with $\lambda = n \max_v p(v)$, such that for any $\epsilon > 0$ there is a metric $d \triangleright \sigma$ and a candidate $a$ in the $(p,q)$-veto core of $\sigma$ with
\[\SC(a,d) \geq \left(1 + \frac{2\lambda}{\m} - \epsilon\right) \SC(\o,d).\]
\end{restatable}
\section{Optimal Distortion with Predictions}
We now provide our second main result, which is a tight analysis of the metric distortion problem through the learning-augmented framework. Specifically, we provide an optimal learning-augmented algorithm for metric distortion and we achieve this by leveraging the results of the previous section. 
    While this demonstrates the applicability of our first main result,
we believe that our results within the learning-augmented framework hold significant independent interest beyond being merely an application of our first result.

In the learning-augmented framework, the regular input that the algorithm receives is augmented with a prediction $\hat{p}$ that may provide the algorithm with some guidance regarding how to approach this instance. For example, in the metric distortion problem, $\hat{p}$ could be a prediction regarding the metric $d$, i.e., all the pairwise distances between agents and voters. Crucially, this prediction can be arbitrarily inaccurate, and the goal is to design algorithms guided by this prediction that guarantee better distortion whenever it happens to be accurate (\emph{consistency}), while maintaining some worst-case distortion guarantee even if the prediction is (arbitrarily) inaccurate (\emph{robustness}). These algorithms receive as input both a preference profile $\sigma$ and prediction $\hat{p}$, and we use $d\triangleright (\sigma, \hat{p})$ to denote the restriction to metrics $d$ for which the prediction $\hat{p}$ is accurate.

Formally, 
the consistency of $\alg$ is the distortion it guarantees when $\hat{p}$ is accurate, 
{\sl i.e.},
\[
\mathsf{consistency}(\alg) = \sup_{\sigma}\ \sup_{d\: :\: d\: \triangleright\: (\sigma,\hat{p})}\ \frac{\SC(\alg(\sigma,\hat{p}),d)}{\SC(\o(d),d)}.
\]
The robustness of $\alg$ is the distortion it guarantees irrespective of the prediction accuracy, {\sl i.e.},
\[
\mathsf{robustness}(\alg) =\sup_{\sigma,~\hat{p}}\ \sup_{d\: :\: d\: \triangleright\: \sigma}\ \frac{\SC(\alg(\sigma,\hat{p}),d)}{\SC(\o(d),d)}. 
\]

In Section~\ref{sec:robustness_consistency_tradeoff}, we consider algorithms augmented with the strongest type of prediction, i.e., a prediction $\hat{p}$ regarding all the pairwise distance among voters and candidates, and we prove an impossibility result regarding the robustness and consistency guarantees that any such algorithm can achieve. Then, in Section~\ref{sec:laalgorithm}, we consider algorithms augmented with a much weaker prediction: a prediction regarding who the optimal candidate is. We propose a family of $(p,q)$-algorithms that use the prediction to determine the $q$ vector, and prove that the robustness-consistency trade-off these algorithms achieve exactly matches the impossibility result shown in Section~\ref{sec:robustness_consistency_tradeoff}. Therefore, despite the fact that they are provided with a weaker type of prediction, these algorithms achieve the optimal distortion and consistency that algorithms with the strongest prediction could achieve.
Finally, we provide a more refined analysis on the distortion as a function of the prediction accuracy (Section~\ref{sec:error}), and as a function of the voters' decisiveness (Section~\ref{sec:decisiveness}).
\subsection{Impossibility Result for Algorithms with Full-Metric Predictions}
\label{sec:robustness_consistency_tradeoff}
We now show that even if an algorithm is augmented with the strongest type of prediction (a prediction regarding all the pairwise distances), it cannot simultaneously achieve $\frac{3-\delta}{1+\delta}$-consistency and better than $\frac{3+\delta}{1-\delta}$ robustness for any $\delta \in \left[0,1\right)$.

\begin{theorem}\label{thm:mainlowerbound}
For any $\delta\in\left[0,1\right)$, let $\alg$ be a deterministic algorithm augmented with a prediction regarding the whole metric that is $\frac{3-\delta}{1+\delta}$-consistent.  Then, for any $\beta < \frac{3+\delta}{1-\delta}$, $\alg$ is not $\beta$-robust. 
\end{theorem}
\begin{proof}
Let  $r(n,\epsilon) \vcentcolon=  (3+\delta - \frac{4}{n} - 2\epsilon)/(1-\delta + \frac{4}{n})$, where $\delta$ and $\beta$ are as in the theorem statement. 
Note that for every $n,\epsilon > 0$ we have $r(n,\epsilon) < \frac{3+\delta}{1-\delta}$. Since this is function is continuous in both $n$ and $\epsilon$, and $\lim_{n \rightarrow \infty, \epsilon \rightarrow 0^+} r(n,\epsilon) = \frac{3+\delta}{1-\delta}$, then there exist $n,\epsilon$ for which $\beta < r(n,\epsilon) < \frac{3+\delta}{1-\delta}$. Fix any $n,\epsilon$ that satisfy this, and also satisfy $1/n - \epsilon > 0$ (since $1/n>0$, we can always choose a small enough value for $\epsilon$ that satisfies both inequalities).

Consider the instance $\mathcal{I}_1=(\V,\C,\d)$ consisting of two candidates $a,b$ whose distance from one another is $2-\epsilon$.  The voters are located in the metric space such that 
$\lceil\frac{1-\delta}{2}n+1\rceil$
of them are placed on $b$, and the rest (of which there are 
$\lfloor\frac{1+\delta}{2}n-1\rfloor$
are placed almost at the halfway point with a slight inclination towards $a$, such that their distance from $b$ is 1 and their distance from $a$ is $1-\epsilon$.  

 \begin{figure}[ht]
 \begin{center}
       \includegraphics[scale=0.4]{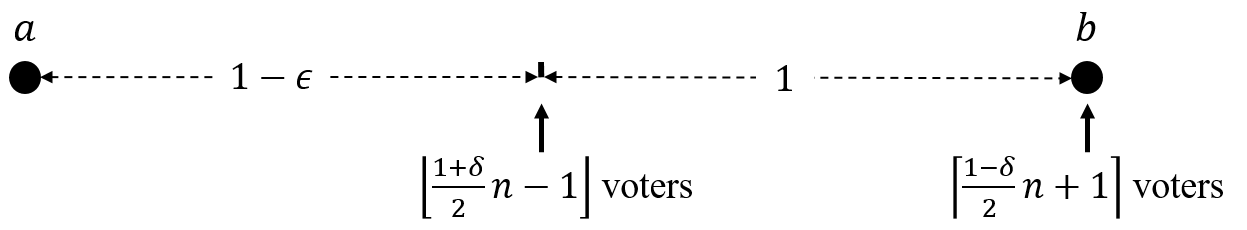}
 \end{center}
\end{figure}

Note that $b$ is the optimal candidate, and $a$ achieves distortion equal to:
\begin{align*}
\frac{\SC(a,d)}{\SC(b,d)}   &= \frac{\lfloor\frac{1+\delta}{2}n-1\rfloor(1-\epsilon) + \lceil\frac{1-\delta}{2}n+1\rceil(2-\epsilon)}{\lfloor\frac{1+\delta}{2}n-1\rfloor\cdot 1} \\
    &=  \frac{\left(\lfloor\frac{1+\delta}{2}n-1\rfloor + \lceil\frac{1-\delta}{2}n+1\rceil \right)(1-\epsilon) +
        \lceil\frac{1-\delta}{2}n+1\rceil \cdot 1}{\lfloor\frac{1+\delta}{2}n-1\rfloor} \\
    &=  \frac{n(1-\epsilon) +
        \lceil\frac{1-\delta}{2}n+1\rceil \cdot 1}{\lfloor\frac{1+\delta}{2}n-1\rfloor} 
    ~\geq~ \frac{n(1-\epsilon) +
        \left(\frac{1-\delta}{2}n+1\right)}{\frac{1+\delta}{2}n} \\ 
    &= \frac{(3-\delta) \frac{n}{2} + 1 -\epsilon n }{\left(1+\delta\right) \frac{n}{2}} 
    ~=~ \frac{3-\delta +\frac{2}{n} -2\epsilon}{1+\delta}
    ~>~ \frac{3-\delta}{1+\delta}
\end{align*}

Therefore, if $\alg$ is provided as input the preference profile $\sigma$ induced by $(\V,\C,\d)$ and the prediction provided to the mechanism is accurate (i.e., it accurately predicts $(\V,\C,\d)$), then to achieve a consistency of $\frac{3-\delta}{1+\delta}$, $\alg$ must output $b$.

Now, consider the instance $\mathcal{I}_2= (\V,\C,\d')$, with the same sets of voters and candidates, but the
$\lceil\frac{1-\delta}{2}n+1\rceil$
voters previously located on $b$ are now located almost at the halfway point between $a$ and $b$ with a slight inclination towards $b$ such that their distance from $b$ is $1-\epsilon$ and their distance from $a$ is 1.  
The rest of the voters (of which there are $\lfloor\frac{1+\delta}{2}n-1\rfloor$) are located on $a$.  
\begin{figure}[ht]  
\begin{center}
   \includegraphics[scale=0.4]{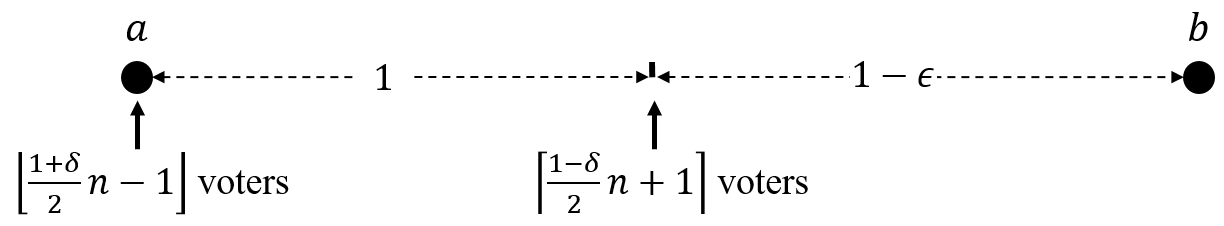} 
\end{center}
\end{figure}
Note that instance $\mathcal{I}_2$ induces the same preference profile $\sigma$ as $\mathcal{I}_1$. Thus, if the real instance is $\mathcal{I}_2$ but $\alg$ received the, now inaccurate, prediction $(\V,\C,\d)$, along with $\sigma$, it would still choose $b$.
Therefore, $\alg$ is not $\beta$-robust, since here $a$ is the optimal candidate, and the output $b$ achieves distortion equal to: 
{\allowdisplaybreaks
\begin{align*}
    \frac{\SC(b,d)}{\SC(a,d)}   &= \frac{\lceil\frac{1-\delta}{2}n+1\rceil(1-\epsilon) + \lfloor\frac{1+\delta}{2}n-1\rfloor(2-\epsilon)}{\lceil\frac{1-\delta}{2}n+1\rceil\cdot 1} \\
        &=      \frac{n(1-\epsilon) + \lfloor\frac{1+\delta}{2}n-1\rfloor\cdot 1}{\lceil\frac{1-\delta}{2}n+1\rceil} 
        ~\geq~  \frac{n(1-\epsilon) + \left(\frac{1+\delta}{2}n-2\right)}{\frac{1-\delta}{2}n+2} \\
        &=      \frac{(3+\delta)\frac{n}{2} - 2 - \epsilon n}{\frac{1-\delta}{2}n+2} 
        ~=~      \frac{3+\delta - \frac{4}{n} - 2\epsilon}{1-\delta + \frac{4}{n}} 
        ~=~ r(n,\epsilon) 
        ~>~ \beta. \qedhere
\end{align*}}
\end{proof}
\subsection{$(p,q)$-Algorithms with Prediction Regarding the Optimal Candidate}\label{sec:laalgorithm}

We now present a family of learning-augmented algorithms, termed $\SimVeto$, which is parameterized by $\delta \in\left[0,1\right)$. Instead of a prediction regarding the whole metric, these algorithms are provided only with a prediction regarding who the optimal candidate is, i.e., some $\hat{p}\in \C$. Note that this is a strictly weaker prediction: given an accurate prediction regarding the whole metric, one can readily infer a prediction regarding who the optimal candidate would be. More importantly, note that the algorithm needs to satisfy the consistency guarantee only on instances where $d \triangleright (\sigma,\hat{p})$. Therefore, if the prediction is regarding the whole metric, then the consistency guarantee binds only if the actual metric, $d$, is exactly the predicted metric. On the other hand, if the prediction is regarding the optimal candidate, there is an infinite number of different metrics $d$ such that $c^*(d)=\hat{p}$, and the algorithm needs to achieve the consistency guarantee for all of them. Nevertheless, despite the weaker prediction that they are provided with, we prove that our algorithms achieve the optimal robustness-consistency trade-off that even algorithms with full-metric predictions can achieve.

\begin{definition}[$\SimVeto$]\label{def:algorithm} Given a $\delta \in (0,1]$, a predicted optimal candidate $\p\in \C$,  and a preference profile $\sigma$, the class of $\SimVeto$ algorithms contains any $(\unip ,\hat{q})$-algorithm with $\hat{q}(\p) = \frac{1-\delta}{1+\delta} \frac{\plu(\p) + \frac{2\delta n}{1-\delta}}{n}$ as the weight of the predicted candidate and $\hat{q}(c) =\frac{1-\delta}{1+\delta} \frac{\plu(c)}{n}$ the weight of any other candidate $c$.

\end{definition}
We refer to $\delta$ as the \emph{confidence parameter} that the designer can choose, depending on how much she trusts the prediction (or, depending on the robustness-consistency trade-off that she prefers). The next theorem shows that for each $\delta$, any $\SimVeto$ algorithm achieves a different Pareto optimal trade-off between robustness and consistency, matching the impossibility result of Theorem~\ref{thm:mainlowerbound}.
\begin{theorem}
\label{thm:optimal_algorithm}
    For any $\delta \in \left[0,1\right)$, $\SimVeto$ achieves $\frac{3-\delta}{1+\delta}$-consistency and $\frac{3+\delta}{1-\delta}$-robustness.
\end{theorem}
\begin{proof}
    We first show that $\SimVeto$ has a distortion at most $\frac{3+\delta}{1-\delta}$ directly from the distortion bound we obtain from Section~\ref{sec:unifromp_generalq}. In other words, any candidate that admits a fractional perfect matching with $(\unip, \hat{q})$ has distortion $\frac{3+\delta}{1-\delta}$. This implies that the \emph{robustness} of $\SimVeto$ algorithm is at most $\frac{3+\delta}{1-\delta}$.
    First note that given some $\delta$ instance $\sigma$, and $\hat{q}$ defined by Definition~\ref{def:algorithm}, we have:
\[\mu(\sigma, \hat{q}) = \min_{c \in \C}\frac{\hat{q}(c)\cdot n}{\plu(c)} = \frac{\frac{1-\delta}{1+\delta}\cdot \frac{\plu(c)}{n} \cdot n}{\plu(c)} =  \frac{1-\delta}{1+\delta}.\]
 From Theorem~\ref{thm:unip_genq}, the distortion of candidate $a$ is at most $1+\frac{2}{\mu}$ by letting $\mu =\frac{1-\delta}{1+\delta}$, we get that the distortion of $(\unip, \hat{q})$-algorithm is at most 
\[1+\frac{2}{\mu} = 1+\frac{2(1+\delta)}{1-\delta} = \frac{3+\delta}{1-\delta}.\]

We now turn to the consistency of $\SimVeto$, i.e., the case $\p = \o$. Assume $a$ is some candidate chosen by a $\SimVeto$ algorithm. The social cost of candidate $a$ is:
{\allowdisplaybreaks
\begin{align*}
    \SC(a,d) &= \sum_{v \in \V}d(v,a)\\
        & = \sum_{v \in \V} n \cdot \unip(v) \cdot d(v,a) \tag{by Def.\  of $\unip$}\\
        & = n \sum_{v \in \V} \sum_{c: (v,c) \in E_a} w((v,c)) \cdot d(v,a) \tag{by Def.\  \ref{def:fractionalperfectmatching}}\\
        & \leq n \sum_{v \in \V} \sum_{c: (v,c) \in E_a} w((v,c)) \cdot d(v,c) \tag{by Def.\  \ref{def:pqgraph}}\\
        & \leq n \sum_{v \in \V} \sum_{c: (v,c) \in E_a} w((v,c)) \cdot \left(d(v,\o) + d(\o, c)\right) \tag{triangle inequality}\\
        & \leq \SC(\o,d) + n \sum_{v \in \V} \sum_{c: (v,c) \in E_a} w((v,c)) \cdot d(\o, c) \\
        & = \SC(\o,d) + n \sum_{c\in \C} \sum_{v \in \V: (v,c) \in E_a} w((v,c)) \cdot d(\o,c)\\
        & =   \SC(\o,d) + n \sum_{c\in \C} \hat{q}(c) \cdot d(\o,c) \tag{by Def.\  \ref{def:fractionalperfectmatching}}\\
        & =  \SC(\o,d) + n \cdot \frac{1-\delta}{1+\delta}  \left(\sum_{c \neq \p} \frac{\plu(c)}{n} \cdot d(\o,c) +  \frac{\plu(\p)+\frac{2\delta n }{1-\delta}}{n} \cdot d(\o,\p)\right)\tag{by Def.\ \ref{def:algorithm}}\\
        & = \SC(\o,d) + \frac{1-\delta}{1+\delta} \sum_{c \neq \o} \plu(c) \cdot d(\o,c) \tag{since $\p = \o$}\\
        & = \SC(\o,d) +\frac{1-\delta}{1+\delta}  \sum_{c\in \C} \sum_{v: \topc(v) = c} d(\o, c)\\
        & \leq \SC(\o,d) + \frac{1-\delta}{1+\delta} \sum_{c\in \C} \sum_{v: \topc(v) = c} (d(c,v) + d(\o,v)) \tag{triangle inequality}\\
        & \leq  \SC(\o,d) + \frac{1-\delta}{1+\delta} \sum_{c}\sum_{v: \topc(v) = c} 2d(\o,v) \tag{since $c= \topc(v) \pref_v \o$}\\
        & \leq \SC(\o,d) + \frac{2(1-\delta)}{1+\delta}\SC(\o,d)
    \leq \frac{3-\delta}{1+\delta}\SC(\o,d). \qedhere
\end{align*}}
\end{proof}

 \subsection{Distortion Bounds Parameterized by Prediction Quality}
\label{sec:error}

Consistency and robustness bounds focus on two extreme cases: the case where the prediction is perfect (consistency) and the case where it can be arbitrarily bad (robustness). To develop a more refined understanding of the distortion that our family of $\SimVeto$ algorithms achieves as a function of the prediction quality, we now seek to prove bounds parameterized by the error of the prediction. 

We first define a natural measure of error, which is proportional to the distance between the prediction, $\p$, and the optimal candidate, $\o$, {\sl i.e.}, $d(\p, \o)$. Note that using this distance alone as a measure of error would not accurately capture the quality of the prediction, since this does not measure how significant this distance is relative to the average cost that each voter would suffer in the optimal solution. For example, if $d(\p, \o)=100$, and the average distance in the optimal solution is $10000$, then the prediction was quite accurate, whereas the prediction would be inaccurate if we scaled the same instance down so that the average distance is $1$. Therefore, to accurately capture the relative quality of the prediction we quantify the error $\error$ of the prediction by normalizing its distance from $\o$ by the average optimal distance, $\SC(\o)/n$, {\sl i.e.},
$$\error\vcentcolon =\frac{n d(\p, \o) }{\SC(\o)}$$ 

\begin{figure}[h]
    \centering
    \includegraphics[width = 0.5 \textwidth]{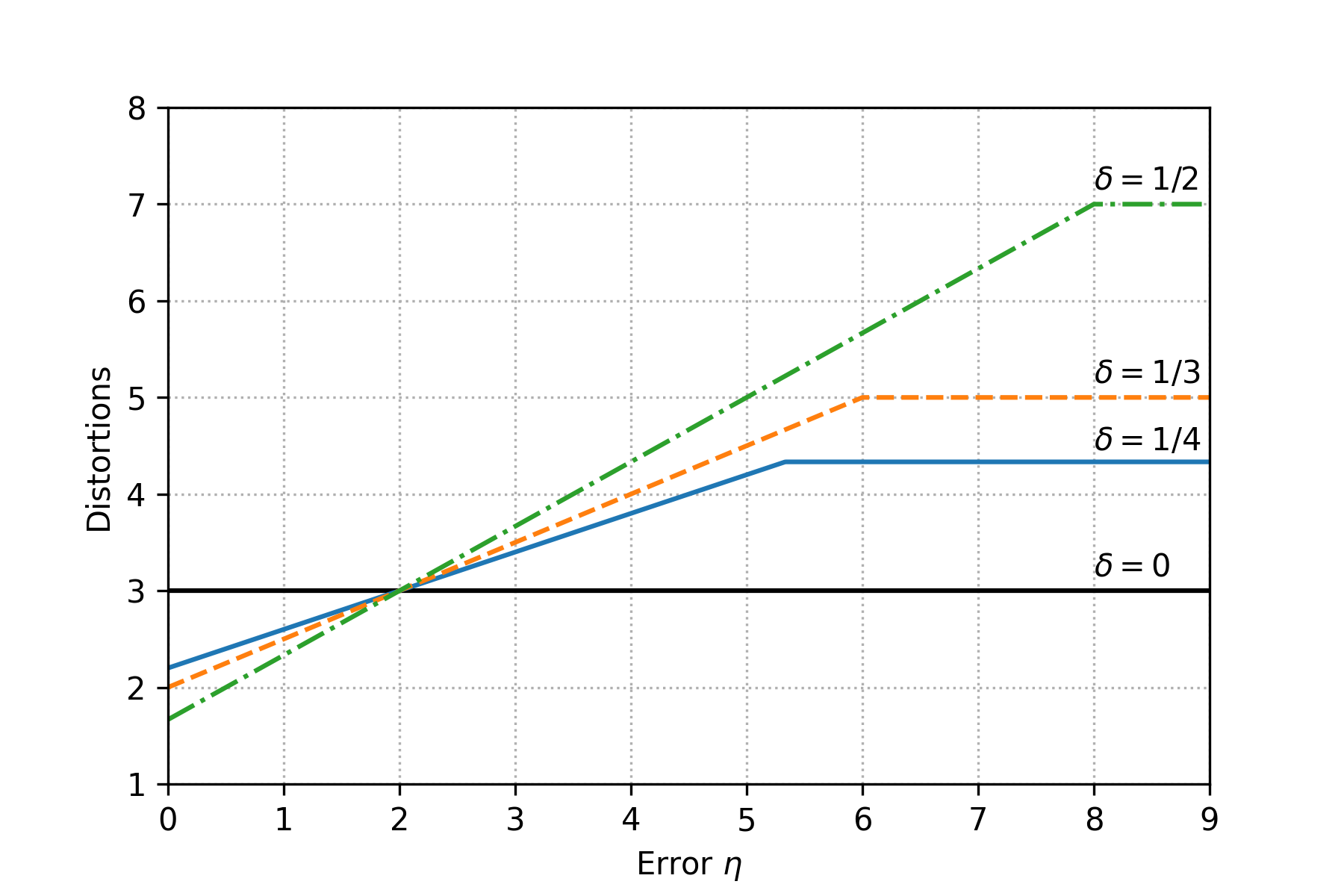}
    \caption{A plot of the distortion upper bound guaranteed by $\SimVeto$ as a function of the prediction error $\error$ for different values of $\delta\in [0,1)$.}
    \label{fig:errordistortion}
\end{figure}

Our main result in this section shows that for each choice of $\delta\in [0, 1)$, our $\SimVeto$ algorithm guarantees distortion at most $\frac{3-\delta+2\delta \error}{1+\delta}$ when the error of the prediction is $\error$. This bound is equal to our consistency bound when the prediction is correct, {\sl i.e.}, $\error=0$, and then grows linearly as a function of $\error$, with a slope that is increasing in $\delta$. This makes sense, as it captures the intuition that trusting the prediction more ({\sl i.e.}, higher values of $\delta$) translates into a worse dependence on the error. Note, however, that the distortion never exceeds our robustness bound, so it gracefully transitions from the consistency bound to the robustness bound as a function of the error. Figure~\ref{fig:errordistortion} demonstrates this bound for different values of $\delta$. For example, as long as $\error \leq 2$, the distortion is at most $3$, for any value of $\delta$.. We defer the proof of the following Proposition to Appendix~\ref{app:error} due to space limitation.

\begin{restatable}{rPro}{properror}\label{prop:error}
For any $\delta \in \left[0,1 \right)$, the distortion achieved by $\SimVeto$ on instances where the prediction $\p$ has error $\error = n \d(\p,\o) / \SC(\o)$ is at most
        $$\min \left\{\frac{3-\delta+2\delta \error}{1+\delta},~ \frac{3+\delta}{1-\delta}\right\}.$$
\end{restatable}
\subsection{Bounds for Metric Spaces with Decisive Voters}\label{sec:decisiveness}
We now refine the achieved distortion bounds even further, using the $\alpha$-decisiveness framework of \citep{AP17}. For $\alpha \in [0,1]$, A voter $v$ is $\alpha$-decisive if $d(v, \topc(v)) \leq \alpha \cdot d(v,c)$, for all $c \neq \topc(v)$. In other words, the distance of voter $v$ from her favorite candidate is at most $\alpha$ times her distance from any other candidate. 
We say 
an instance $(\V,\C,\d)$
is $\alpha$-decisive if every voter is $\alpha$-decisive. Note that when $\alpha = 0$, the distance between every voter and her favorite candidate is $0$. This captures the \emph{peer selection} setting, in which the set of voters is the set of candidates, and each voter ranks herself first. Therefore, the positive results presented in the section for $\alpha = 0$ apply to this setting as well. Since the proofs of the decisiveness results are very similar to some of our earlier proofs, we defer them to Appendix~\ref{app:decisiveness} and directly provide the most general bound that is parameterized both by the decisiveness parameter $\alpha$, and the error parameter $\error$.

\begin{proposition}\label{prop:error-devisiveness}
For any $\delta \in \left[0,1 \right)$ and $\alpha \in [0,1]$, the distortion achieved by $\SimVeto$ on $\alpha$-decisive instances where the prediction $\p$ has error $\error = n \d(\p,\o) / \SC(\o)$
is at most
        $$\frac{2 + \alpha -\alpha \delta}{1+\delta} + \min \left\{\frac{2\delta \error}{1+\delta},~ \frac{8\delta}{(1+\delta)(1-\delta)}\right\}.$$
\end{proposition}

\begin{figure}[h]
    \centering
    \includegraphics[width = 0.48 \textwidth]{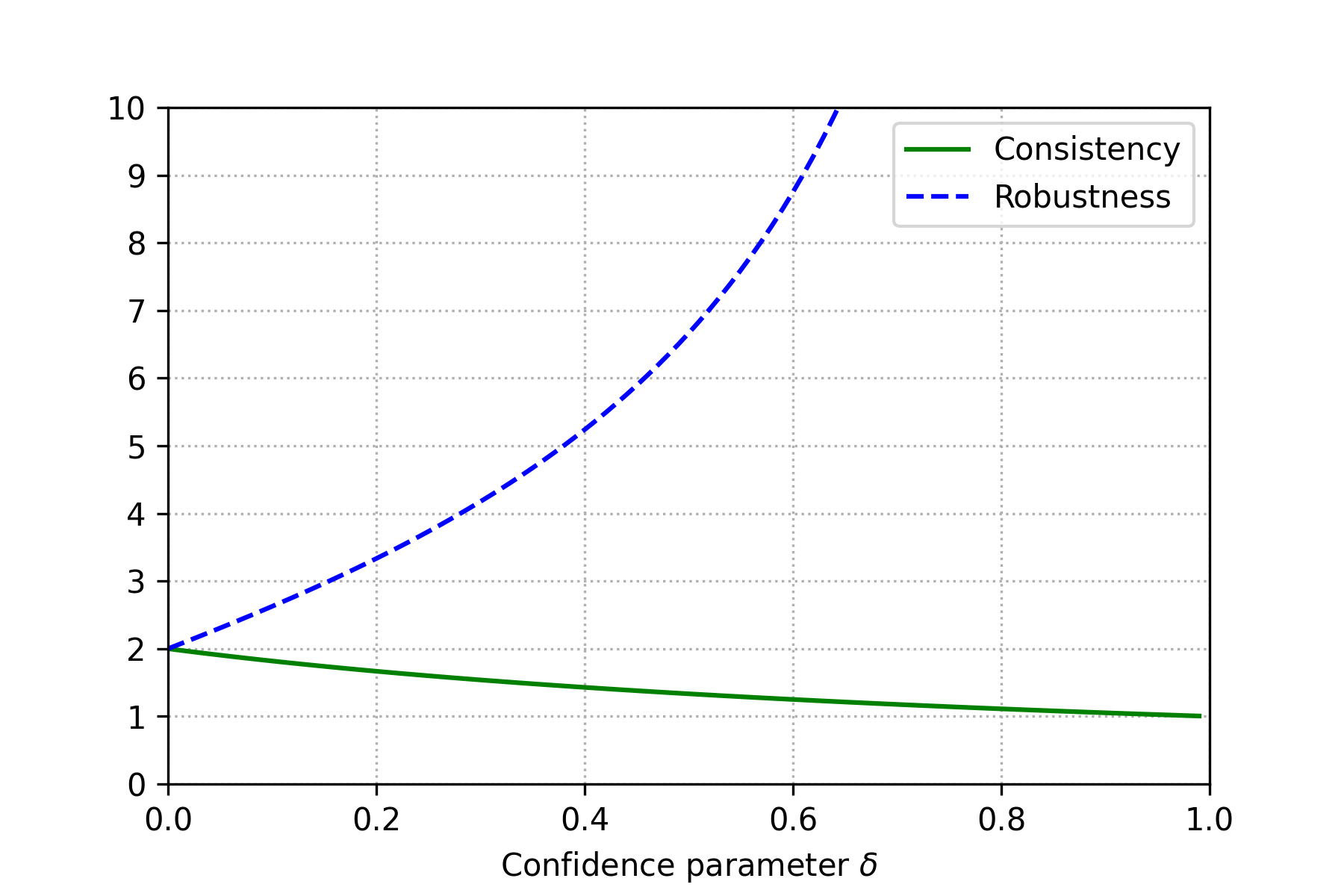}
    \includegraphics[width = 0.48 \textwidth]{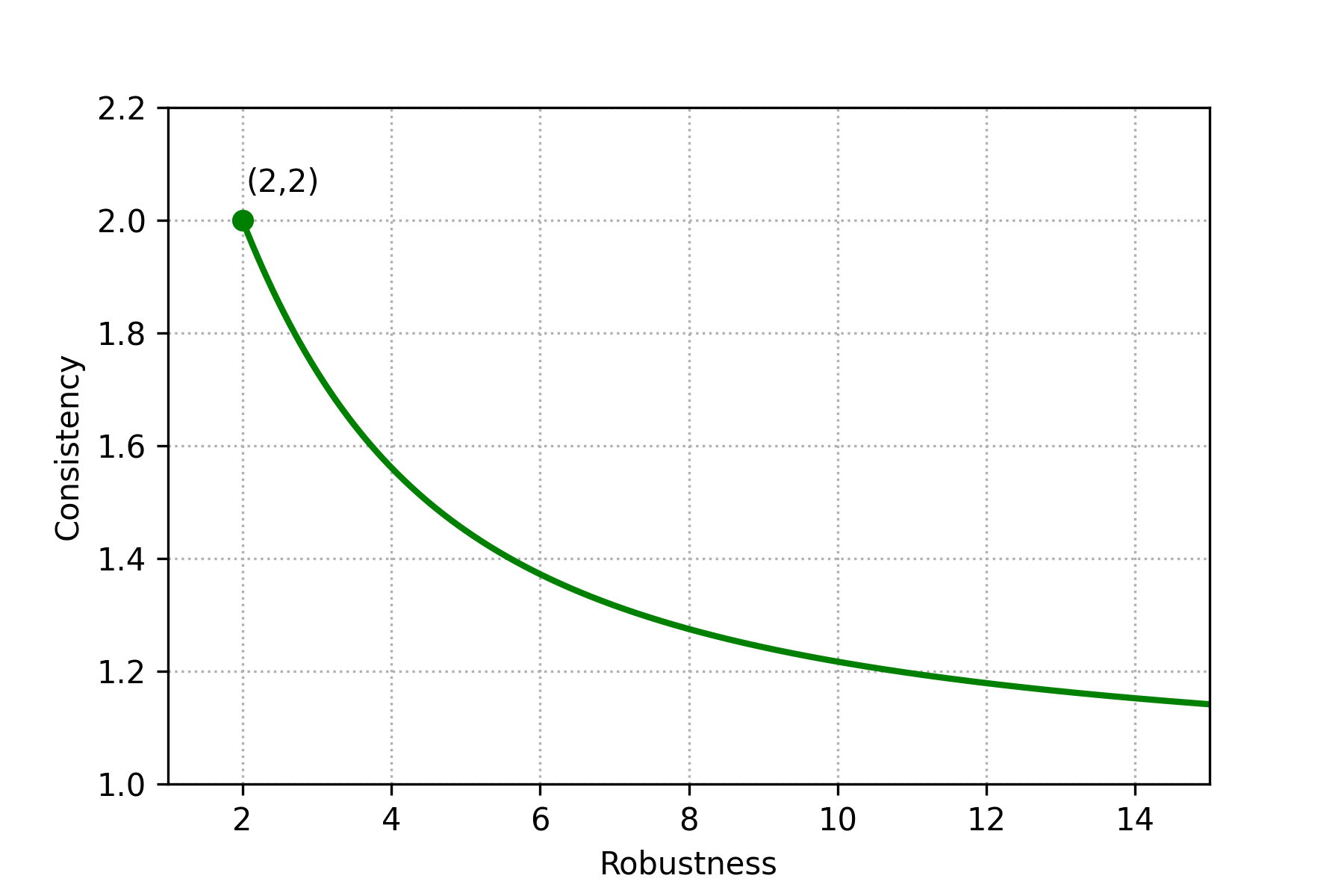}
    \caption{The analogue of Figure \ref{fig:optimal_trade-off} for $0$-decisive instances (peer selection setting).
    On the left, a plot showing the consistency and robustness bounds achieved by $\SimVeto$ as a function of the confidence parameter $\delta$. On the right, a plot capturing the 
    consistency-robustness pairs that can be achieved by $\SimVeto$ for 
    any chosen $\delta \in [0,1)$.}
    \label{fig:decisive_optimal_tradeoff}
\end{figure}

If we focus our attention on the interesting case where $\alpha=0$, {\sl i.e.}, the peer selection case, then this implies the following bound.
\begin{corollary}\label{cor:error-devisiveness=0}
For any $\delta \in \left[0,1 \right)$ and $\alpha=0$, the distortion achieved by $\SimVeto$ on $\alpha$-decisive instances where the prediction $\p$ has error $\error = n \d(\p,\o) / \SC(\o)$ 
is at most
        $$\frac{2}{1+\delta} + \min \left\{\frac{2\delta \error}{1+\delta},~ \frac{8\delta}{(1+\delta)(1-\delta)}\right\}.$$
\end{corollary}

In the metric distortion problem with 0-decisive voters (the peer selection setting),
the best known achievable distortion is 2. Our algorithm recovers this guarantee 
for $\delta = 0$. For $\delta>0$,
our algorithm achieves a consistency strictly better than 2, at the cost of higher robustness --- Figure~\ref{fig:decisive_optimal_tradeoff} illustrates the achieved consistency-robustness pairs obtained by varying $\delta$. 
Note that as long as the value of $\eta$ is at most 1, our algorithm ensures that the achieved distortion remains at most 2.

\section{Conclusion and Future Work}
Our contribution in this work is twofold: \emph{i)} we establish an new distortion upper bound for candidates om the $(p,q)$-veto core for general $p$ and $q$ and \emph{ii)} we initiate the study of distortion problems through the learning-augmented framework, leading to an optimal learning-augmented algorithm guided by a prediction for the optimal candidate.

An interesting avenue for future research involves a deeper exploration of the implications of our distortion bounds for the $(p,q)$-veto core. 
A particularly intriguing problem within this direction is identifying other well-motivated alternatives for the $q$ vector. 
For example, \citet{KK23} propose using $k$-plurality scores of candidates (the number of times they appear among the top $k$ choices of a voter) to generate a $q$, which interpolates between $q^\plu$ (for $k=1$) and uniform $q$ (for $k=m$). 
Another interesting direction is to examine the potential for alternative parameterizations of the distortion bounds. Identifying even more refined upper bounds as a function of $p$, $q$ and $\plu$, or as a function of other parameters.

Our results also pave the way for revisiting many other interesting problems within the distortion literature using the learning-augmented framework (see \cite{AFSV21} for a recent survey on the distortion literature). 
Examples include the multi-winner variant of the metric distortion problem 
\cite{CSV22}, the distortion of one-sided matching \cite{filos2014truthful}, as well as
the utilitarian distortion setting with a single winner \cite{procaccia2006distortion} or multiple winners \cite{caragiannis2017subset}. Another natural direction is to apply this framework to randomized metric distortion, especially given the recent randomized algorithm achieving distortion better than $3$~\cite{CWRW24}. Can this be extended toward more general improvements on the robustness-consistency Pareto frontier achievable via randomization? 
\section*{Acknowledgement}
We would like to thank the anonymous reviewers for their detailed feedback and comments that helped significantly improve the exposition of this paper.

\bibliographystyle{plainnat}
\bibliography{bibliography,biblio}
\appendix
\section{Missing Proofs from Section~\ref{sec:generalp_generalq}}
\label{app:pqreduction}

\lemgenpunipreduction*
\begin{proof}
We let $p_m = \max_v p(v)$, and for each voter $v$ we define $\lambda_v = p_m - p(v)$. We note that $0 \leq \lambda_v \leq 1$. 
We construct the following $q'$ vector as a function of the initial $p,q$ as follows: for any candidate $c \in \C$, we define:
\[q'(c) = \frac{1}{\lambda} \cdot \left(q(c)+ \sum_{v: (v,c) \in E_a} \frac{w((v,c))\cdot \lambda_v}{p(v)} \right).\]
We first show that $q' \in \Delta(\C)$. To this end, consider:
{\allowdisplaybreaks
\begin{align*}
    \sum_{c}q(c') & = \frac{1}{\lambda} \cdot \sum_{c} \left(q(c)+ \sum_{v: (v,c) \in E_a} \frac{w((v,c))\cdot \lambda_v}{p(v)}\right)\\
     & =  \frac{1}{\lambda} \left( \sum_{c} q(c) + \sum_{c}\sum_{v: (v,c) \in E_a} \frac{w((v,c))\cdot \lambda_v}{p(v)}\right)\\
     & = \frac{1}{\lambda} \cdot \left( 1 + \sum_{c}\sum_{v: (v,c) \in E_a} \frac{w((v,c))\cdot \lambda_v}{p(v)}\right) \tag{since $q \in \Delta(c)$}\\
     & = \frac{1}{\lambda} \cdot \left( 1 + \sum_{v}\sum_{c: (v,c) \in E_a} \frac{w((v,c))\cdot \lambda_v}{p(v)}\right) \\
     & = \frac{1}{\lambda} \cdot \left( 1 + \sum_{v} \lambda_v\right) \tag{by Def.\ ~\ref{def:fractionalperfectmatching}}\\
     & = \frac{1}{\lambda} \cdot (\lambda) ~=~ 1 \tag{by Def.\  of $\lambda$}
\end{align*}}

We now show that $G^\sigma_{\unip,q'}(a)$ admits a fractional perfect matching. Consider the following weight function $w'$: for any $(v,c) \in E_a$, we let:
\[w'((v,c)) = \frac{1}{\lambda} \cdot \left(1 + \frac{\lambda_v}{p(v)} \right) \cdot w((v,c)).\]
First, consider any candidate $c$:
\begin{align*}
    \sum_{v:(v,c) \in E_a} w'((v,c)) & = \frac{1}{\lambda} \cdot \sum_{v: (v,c) \in E_a}  \left(1 + \frac{\lambda_v}{p(v)} \right) \cdot w((v,c))\\
    & = \frac{1}{\lambda} \cdot \sum_{v: (v,c) \in E_a} \left( w((v,c))+ \frac{ w((v,c)) \cdot \lambda_v}{p(v)} \right)\\
    & = \frac{1}{\lambda} \cdot \left(q(c) + \sum_{v: (v,c) \in E_a} \frac{ w((v,c)) \cdot \lambda_v}{p(v)} \right) \tag{by Def.\ ~\ref{def:fractionalperfectmatching}}\\
    & = q'(c)
\end{align*}
Now, consider any voter $v$:
\begin{align*}
    \sum_{c: (v,c) \in E_a} w'((v,c)) & = \frac{1}{\lambda} \cdot \left(1 + \frac{\lambda_v}{p(v)} \right)\cdot \sum_{c \in \C} w((v,c))\\
                            &= \frac{1}{\lambda} \cdot \left(1 + \frac{\lambda_v}{p(v)} \right) \cdot p(v)\\
                            & = \frac{p(v)+\lambda_v}{\lambda}\\
                            & = \frac{p_m}{\sum_{v}(p(v) + \lambda_v)} \tag{by Def.\  of $\lambda$ and $p\in \Delta(\V)$}\\
                        & = \frac{p_m}{n\cdot p_m} ~=~ \frac{1}{n} ~=~ \unip(v). 
\end{align*}

Finally, we consider the change in $\mu$ value from $q$ to $q'$. First, since $\lambda_v \geq 0$, we have
\[q'(c) = \frac{1}{\lambda} \cdot \left(q(c)+ \sum_{v: (v,c) \in E_a} \frac{w((v,c))\cdot \lambda_v}{p(v)} \right) \geq \frac{q(c)}{\lambda}.\]
Therefore, for any candidate $c$, we have:
\[\frac{q'(c) \cdot n}{\plu(c)} \geq \frac{1}{\lambda} \cdot \frac{q(c) \cdot n}{\plu(c)}.\]
Since the pointwise change of any candidate $c$ is bounded by $\frac{1}{\lambda}$, we therefore have 
\[\mu(\sigma, q') = \min_{c \in \C} \frac{q'(c) \cdot n}{\plu(c)} \geq \min_{c \in \C} \frac{1}{\lambda} \cdot \frac{q(c) \cdot n}{\plu(c)} = \frac{\mu(\sigma, q)}{\lambda}. \qedhere \]
\end{proof}

\thmpqlowerbound*
\begin{proof}
Consider the following instance ($\V$,$\C$,d) consisting of two candidates $a$, $b$ below.
\begin{figure}[ht]  
\begin{center}
    \includegraphics[scale=0.5]{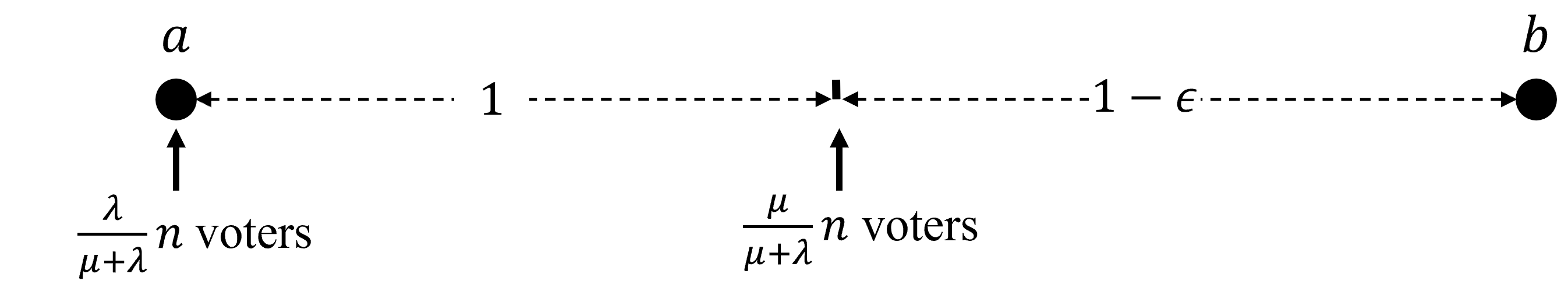}
\end{center}
\end{figure}
We first note that in this instance, $\plu(a) = \frac{\lambda}{\m+\lambda}n$ and $\plu(b) = \frac{\m}{\m+\lambda}n$. Now consider $q$ such that $q(a) =\frac{\m\lambda}{\m+\lambda}$ and $q(b) = \frac{\m + \lambda - \m\lambda}{\m+\lambda}$, it is easy to verify that $\min_c \frac{q(c)n}{\plu(c)} = \frac{q(a)n}{\plu(a)}= \m$. We now partition the voters into $V_1$ and $V_2$, where $V_1$ is the set of voters co-located with candidate $a$ and $V_2$ is the set of voters co-located in the midpoint. For any $v \in V_1$, we set $p(v) = \frac{1}{n} - \frac{(\lambda-1)\mu}{\lambda n}$, and for any voter $v' \in V_2$, we set $p(v') = \frac{1}{n}+\frac{\lambda-1}{n}$. Note that $\max_v(p(v)) = \frac{1}{n}+\frac{\lambda-1}{n}$ and $n \cdot \max_v(p(v)) = 1+\lambda-1 = \lambda$.
\[\]
\begin{figure}
	\begin{subfigure}[b]{0.5\textwidth}
		\centering
		\[\begin{array}{ll}
			  v \in V_1 :& a \succ b  \quad |V_1| = \frac{\lambda}{\mu+\lambda}\\
			  v \in V_2 :& b \succ a  \quad |V_2| = \frac{\mu}{\mu+\lambda}
		\end{array}\]
		\caption{Preference Profile $\sigma$}
		\label{fig:rankings_lowerbound}
	\end{subfigure}
	\begin{subfigure}[b]{0.5\textwidth}
		\centering
		\begin{tikzpicture}[scale=.75]
			\tikzstyle{main_node} = [circle,fill=white,draw,minimum size=1.2em,inner sep=0pt]

			\node[main_node] (1) at (0, 0) {$V_1$};
			\node[main_node] (2) at (0, -1) {$V_2$};
			\node[main_node] (a) at (2, 0) {$a$};
			\node[main_node] (b) at (2, -1) {$b$};

			\draw (1) node [left=.5em] {\tiny $p(v) = \frac{1}{n} - \frac{(\lambda-1)\mu}{\lambda n}$};
			\draw (2) node [left=.5em] {\tiny $p(v) = \frac{1}{n} + \frac{\lambda-1}{n}$};

			\draw (a) node [right=.5em] {\tiny $q(a) = \frac{\mu\lambda}{\mu+\lambda}$};
			\draw (b) node [right=.5em] {\tiny $q(b) = \frac{\mu+ \lambda -\mu\lambda}{\mu+\lambda}$};



   		\draw (1)--(b);
            \draw (2)--(b);
			\draw (2)--(a);
			\draw (1)--(b) [draw=red, line width=.75mm];
			\draw (2)--(a) [draw=red, line width=.75mm];
		\end{tikzpicture}
		\caption{Domination Graph $G^\sigma_{p, q}(b)$}
		\label{fig:b-domination}
	\end{subfigure}
 \caption{the preference profile and the domination graph induced from the instance and the choice of $p$, $q$.}
 \end{figure}
We now argue that candidate $b$, given the $p$, $q$ vector defined above has a fractional perfect matching. Consider the graph $G^\sigma_{p,q}(b)$, note that, for voters $v \in V_1$ they the have only have edges to $b$, whereas the voters in $V_2$ has edges to both candidate $a,b$. Consider a simple weight function $w$ where $w((v,b)) = p(v) = \frac{1}{n} - \frac{(\lambda-1)\mu}{\lambda n}$ and $w((v,a)) = 0$ for all $v \in V_1$ and $w((v',a)) = \frac{1}{n}+\frac{\lambda-1}{n}$ and $w((v,b)) =0$ for all $v' \in V_2$. We first note that for each voter $v$, $w((v,a)) + w((v,b)) = p(v)$. Now consider the candidates,
we have:
\[\sum_{v: (v,b)\in E_b} w((v,b)) = \left(\frac{1}{n} - \frac{(\lambda-1)\mu}{\lambda n}\right)\cdot |V_1| = \left(\frac{1}{n} - \frac{(\lambda-1)\mu}{\lambda n}\right)\cdot \frac{\lambda}{\m+\lambda}n =\frac{\m+\lambda-\m\lambda}{\m+\lambda}= q(b).\]
\[\sum_{v: (v,a) \in E_b} w((v,a)) = \left(\frac{1}{n}+\frac{\lambda-1}{n}\right) \cdot |V_2| = \left(\frac{1}{n}+\frac{\lambda-1}{n}\right) \cdot \frac{\m}{\m+\lambda}n= \frac{\m \lambda}{\m+\lambda} = q(a).\]
We therefore conclude that $G^\sigma_{p,q}(b)$ admits a fractional perfect matching. We now compute the distortion of candidate $b$. To this end, we first compute the optimal social cost of the instance, note that the optimal candidate in this case is candidate $a$, we have:
\[\SC(a,d) = \frac{\m}{\m+\lambda}n \cdot 1 + \frac{1}{1+\m}n\cdot 0 =\frac{\m}{\m+\lambda}n.\]
The cost of candidate $b$ is:
\begin{align*}
   \SC(b,d) = \frac{\mu}{\m+\lambda}n \cdot (1-\epsilon) + \frac{\lambda}{\m+\lambda}n \cdot (2-\epsilon)\\
= \frac{\m + 2\lambda}{\mu+\lambda}n - \epsilon n = \left(1+\frac{2\lambda}{\mu} - \frac{\epsilon(\mu+\lambda)}{\mu}\right)\SC(a,d) 
\end{align*}

the distortion converges to $1+\frac{2\lambda}{\m}$ as $\epsilon$ converges to $0$. 
\end{proof}
\section{Missing Proofs from Section~\ref{sec:error}}\label{app:error}
\properror*
\begin{proof}
For any $\delta \in [0,1)$, 
assume that $a$ is the candidate output by a given execution of $\SimVeto$,
when provided with a prediction $\p$ with error $\error=n d(\p, \o) / \SC(\o,d)$.
Theorem \ref{thm:optimal_algorithm} already shows that the worst-case distortion of $\SimVeto$ is at most $\frac{3+\delta}{1-\delta}$ for any $\delta$, so we just need to prove that it is also at most $\frac{3-\delta+2\delta \error}{1+\delta}$.

The social cost of $a$ can be upper bounded as follows:
\allowdisplaybreaks
\begin{align*}
    \SC(a,d) &= \sum_{v \in \V}d(v,a)\\
        & = \sum_{v \in \V} n \cdot \unip(v) \cdot d(v,a) \tag{by Def.\  of $\unip$}\\
        & = n \sum_{v \in \V} \sum_{c: (v,c) \in E_a} w((v,c)) \cdot d(v,a) \tag{by Def.\  \ref{def:fractionalperfectmatching}}\\
        & \leq n \sum_{v \in \V} \sum_{c: (v,c) \in E_a} w((v,c)) \cdot d(v,c) \tag{by Def.\  \ref{def:pqgraph}}\\
        & \leq n \sum_{v \in \V} \sum_{c: (v,c) \in E_a} w((v,c)) \cdot \left(d(v,\o) + d(\o, c)\right) \tag{triangle inequality}\\
        & \leq \SC(\o,d) + n \sum_{v \in \V} \sum_{c: (v,c) \in E_a} w((v,c)) \cdot d(\o, c) \\
        & = \SC(\o,d) + n \sum_{c\in \C} \sum_{v \in \V: (v,c) \in E_a} w((v,c)) \cdot d(\o,c)\\
        & =   \SC(\o,d) + n \sum_{c\in \C} \hat{q}(c) \cdot d(\o,c) \tag{by Def.\  \ref{def:fractionalperfectmatching}}\\
        & =  \SC(\o,d) + n \cdot \frac{1-\delta}{1+\delta}  \left(\sum_{c \neq \p} \frac{\plu(c)}{n} \cdot d(\o,c) +  \frac{\plu(\p)+\frac{2\delta n }{1-\delta}}{n} \cdot d(\o,\p)\right)\tag{by Def.\ ~\ref{def:algorithm}}\\
        & = \SC(\o,d)  +\frac{2\delta \error}{1+\delta}\SC(\o,d) + \frac{1-\delta}{1+\delta} \sum_{c \neq \o} \plu(c) \cdot d(\o,c) \tag{by $d(\o,\p)= \error \cdot \SC(\o,d)/n$}\\
        & = \frac{1+\delta + 2\delta \error}{1+\delta}\SC(\o,d) +\frac{1-\delta}{1+\delta}  \sum_{c\in \C} \sum_{v: \topc(v) = c} d(\o, c)\\
        & \leq \frac{1+\delta + 2\delta \error}{1+\delta}\SC(\o,d) + \frac{1-\delta}{1+\delta} \sum_{c\in \C} \sum_{v: \topc(v) = c} (d(c,v) + d(\o,v)) \tag{triangle inequality}\\
        & \leq  \frac{1+\delta + 2\delta \error}{1+\delta}\SC(\o,d) + \frac{1-\delta}{1+\delta} \sum_{c}\sum_{v: \topc(v) = c} 2d(\o,v) \tag{since $c= \topc(v) \pref_v \o$}\\
        & \leq \frac{1+\delta + 2\delta \error}{1+\delta}\SC(\o,d) + \frac{2(1-\delta)}{1+\delta}\SC(\o,d)
    \leq \left(\frac{3-\delta+2\delta \error}{1+\delta}\right)\SC(\o,d). \qedhere
\end{align*}
\end{proof}

\section{Missing Proofs from Section~\ref{sec:decisiveness}}\label{app:decisiveness}

\subsection{Consistency Bound for $\alpha$-Decisive Instances}
\label{sec:consistencyalpha}
\begin{proposition}
\label{prop:consistencyalpha}
    For any $\delta \in [0,1)$ and $\alpha \in [0,1]$,
    assume that $a$ is the candidate output by a given execution of $\SimVeto$
    on an $\alpha$-decisive instance
    when provided with the correct prediction $\p = \o$.
    Then we have
    $$\SC(a,d) \leq \frac{2+\alpha-\alpha\delta}{1+\delta}\cdot \SC(\o,d).$$
\end{proposition}
\begin{proof}
The social cost of the returned candidate $a$ can be upper bounded as follows:
{\allowdisplaybreaks
\begin{align*}
    \SC(a,d) &= \sum_{v \in \V}d(v,a)\\
        & = \sum_{v \in \V} n \cdot \unip(v) \cdot d(v,a) \tag{by Def.\  of $\unip$}\\
        & = n \sum_{v \in \V} \sum_{c: (v,c) \in E_a} w((v,c)) \cdot d(v,a) \tag{by Def.\  \ref{def:fractionalperfectmatching}}\\
        & \leq n \sum_{v \in \V} \sum_{c: (v,c) \in E_a} w((v,c)) \cdot d(v,c) \tag{by Def.\  \ref{def:pqgraph}}\\
        & \leq n \sum_{v \in \V} \sum_{c: (v,c) \in E_a} w((v,c)) \cdot \left(d(v,\o) + d(\o, c)\right) \tag{triangle inequality}\\
        & \leq \SC(\o,d) + n \sum_{v \in \V} \sum_{c: (v,c) \in E_a} w((v,c)) \cdot d(\o, c) \\
        & = \SC(\o,d) + n \sum_{c\in \C} \sum_{v \in \V: (v,c) \in E_a} w((v,c)) \cdot d(\o,c)\\
        & =   \SC(\o,d) + n \sum_{c\in \C} \hat{q}(c) \cdot d(\o,c) \tag{by Def.\  \ref{def:fractionalperfectmatching}}\\
        & =  \SC(\o,d) + n \cdot \frac{1-\delta}{1+\delta}  \left(\sum_{c \neq \p} \frac{\plu(c)}{n} \cdot d(\o,c) +  \frac{\plu(\p)+\frac{2\delta n }{1-\delta}}{n} \cdot d(\o,\p)\right)\tag{by Def. \ref{def:algorithm}}\\
        & = \SC(\o,d) + \frac{1-\delta}{1+\delta} \sum_{c \neq \o} \plu(c) \cdot d(\o,c) \tag{since $\p = \o$}\\
        & = \SC(\o,d) +\frac{1-\delta}{1+\delta}  \sum_{c\in \C} \sum_{v: \topc(v) = c} d(\o, c)\\
        & \leq \SC(\o,d) + \frac{1-\delta}{1+\delta} \sum_{c\in \C} \sum_{v: \topc(v) = c} (d(c,v) + d(\o,v)) \tag{triangle inequality}\\
        & \leq \SC(\o,d) + \frac{1-\delta}{1+\delta} \sum_{c\in \C} \sum_{v: \topc(v) = c} (\alpha d(\o,v) + d(\o,v)) \tag{$\alpha$-decisiveness}\\
        & \leq  \SC(\o,d) + \frac{1-\delta}{1+\delta} \sum_{c}\sum_{v: \topc(v) = c} (1+\alpha)d(\o,v) \\
        & \leq \SC(\o,d) + \frac{(1+\alpha)(1-\delta)}{1+\delta}\SC(\o,d)
    \leq \frac{2+\alpha-\alpha\delta}{1+\delta}\SC(\o,d). \qedhere
\end{align*}}
\end{proof}

\subsection{Robustness Bound for $\alpha$-Decisive Instances}\label{sec:robustness_predictionalpha}
\begin{proposition}
\label{prop:robustness_mainalpha}
        For any $\delta \in [0,1)$ and $\alpha \in [0,1]$, 
        assume that $\p$ is the candidate output by a given execution of $\SimVeto$ on an $\alpha$-decisive instance,
        {\sl i.e.}, the algorithm returns the prediction.
        Then we have
        $$\SC(\p,d) \leq \left(\frac{2 + \alpha -\alpha \delta}{1+\delta} + \frac{8\delta}{(1+\delta)(1-\delta)}\right) \cdot \SC(\o,d).$$
\end{proposition}
\begin{proof}
We first note that by Definition~\ref{def:pqalgorithm}, any candidates output by $\SimVeto$ is a member of $(p,q)$-veto core for the corresponding $p,q$.The social cost of the returned candidate $a$ can then be upper bounded as follows given a $\alpha$-decisiveness instance. Note that the proof is very similar with the proof of Theorem~\ref{thm:unip_genq}.
 {\allowdisplaybreaks
 \begin{align*}
 \SC(a,d) &= \sum_{v \in \V}d(v,a)\\
        & = \sum_{v \in \V} n \cdot p(v) \cdot d(v,a) \tag{since $p$ is uniform}\\
        & = n \sum_{v \in \V} \sum_{c \in \C: (v,c) \in E_a} w((v,c)) \cdot d(v,a) \tag{by Def.\  \ref{def:fractionalperfectmatching}}\\
        & = n \sum_{c \in \C}\sum_{v: (v,c) \in E_a} \left(\frac{\mu \cdot \plu(c)}{q(c)n} + \left(1-\frac{\mu \cdot \plu(c)}{q(c)n}\right)\right) \cdot w((v,c)) \cdot d(v,a)\\        
        & = n \sum_{c \in \C}\frac{\mu \cdot \plu(c)}{q(c)n}\sum_{v: (v,c) \in E_a} w((v,c)) \cdot d(v,a) + n\sum_{c \in \C}\left(1-\frac{\mu \cdot \plu(c)}{q(c)n}\right)\sum_{v: (v,c) \in E_a} w((v,c)) \cdot d(v,a)\\ 
        & \leq n  \sum_{c \in \C}\frac{\mu \cdot \plu(c)}{q(c)n}\sum_{v: (v,c) \in E_a} w((v,c)) \cdot d(v,c) \tag{by $a \pref_v c$}\\
        & \quad + n\sum_{c \in \C}\left(1-\frac{\mu \cdot \plu(c)}{q(c)n}\right)\sum_{v: (v,c) \in E_a} w((v,c)) \cdot d(v,a) \\
        & \leq n \sum_{c \in \C}\frac{\mu \cdot \plu(c)}{q(c)n}\sum_{v: (v,c) \in E_a} w((v,c)) \cdot (d(v,\o)+d(\o,c)) \\
        &\quad+  n\sum_{c \in \C}\left(1-\frac{\mu \cdot \plu(c)}{q(c)n}\right)\sum_{v: (v,c) \in E_a} w((v,c)) \cdot (d(v,\o)+d(\o,a)) \tag{triangle inequality}\\
        & = n \sum_{c \in \C}\left(\frac{\mu \cdot \plu(c)}{q(c)n} 
    + 1-\frac{\mu \cdot \plu(c)}{q(c)n}\right) \sum_{v: (v,c) \in E_a} w((v,c)) \cdot d(v,\o) \\
        &\quad+ n \sum_{c \in \C}\frac{\mu \cdot \plu(c)}{q(c)n}\sum_{v: (v,c) \in E_a} w((v,c)) \cdot d(\o,c)\\
        &\quad+  n\sum_{c \in \C}\left(1-\frac{\mu \cdot \plu(c)}{q(c)n}\right)\sum_{v: (v,c) \in E_a} w((v,c)) \cdot d(\o,a)\\
        & = n \sum_{v \in \V}\sum_{v: (v,c) \in E_a} w((v,c)) \cdot d(v,\o) \\
        &\quad+ n \sum_{c \in \C}\frac{\mu \cdot \plu(c)}{q(c)n}\sum_{v: (v,c) \in E_a} w((v,c)) \cdot d(\o,c) \tag{rearranging the sum}\\
        &\quad+  n\sum_{c \in \C}\left(1-\frac{\mu \cdot \plu(c)}{q(c)n}\right)\sum_{v: (v,c) \in E_a} w((v,c)) \cdot d(\o,a)\\
        & = \SC(\o,d) + n\sum_{c \in \C}\frac{\mu \cdot \plu(c)}{q(c)n}\sum_{v: (v,c) \in E_a} w((v,c)) \cdot d(\o,c) \tag{by Def.\ ~\ref{def:fractionalperfectmatching}}\\
        &\quad+ n\sum_{c \in \C}\left(1-\frac{\mu \cdot \plu(c)}{q(c)n}\right)\sum_{v: (v,c) \in E_a} w((v,c)) \cdot d(\o,a) \\
        & =  \SC(\o,d) + n\sum_{c \in \C}\frac{\mu \cdot \plu(c)}{q(c)n} \cdot q(c)\cdot d(\o,c)\\
        &\quad +n\sum_{c \in \C}\left(1-\frac{\mu \cdot \plu(c)}{q(c)n}\right)q(c)\cdot d(\o,a) \tag{by Def.\  \ref{def:fractionalperfectmatching}}\\
        & =  \SC(\o,d) + n \sum_{c \in \C}\sum_{v:  \topc(v) = c}\frac{\m}{n} \cdot d(\o,c) + n(1-\m)\cdot d(\o,a) \\
        & =  \SC(\o,d) + \m \sum_{c \in \C}\sum_{v:  \topc(v) = c} (d(\o,v)+d(v,c)) + n(1-\m)\cdot d(\o,a) \tag{triangle inequality}\\
      &\leq \SC(\o,d) + \m \sum_c\sum_{v: \topc(v) = c} (1+\alpha)d(\o,v)+ n(1-\m)d(\o,a) \tag{$\alpha$-decisiveness}\\
       & \leq \SC(\o,d) + (1+\alpha)\m\cdot\SC(\o,d) + n(1-\m) \cdot \frac{2\SC(\o,d)(1+\m)}{\m n} \tag{by Lemma~\ref{lem:opt_bound_matchinganyq}}\\
       & \leq \left(1+(1+\alpha)\m+\frac{2(1-\m)(1+\m)}{\m}\right)\SC(\o,d)\\
       &=\left(\frac{2 + \alpha -\alpha \delta}{1+\delta} + \frac{8\delta}{(1+\delta)(1-\delta)}\right) \cdot \SC(\o,d). \tag{since $\mu = \frac{1-\delta}{1+\delta}$}
        \end{align*}}
    This concludes the proof.
 \end{proof}

\subsection{Bounds Parameterized by Prediction Quality for $\alpha$-Decisive Instances}
\label{sec:erroralpha}

\begin{proposition}
\label{prop:erroralpa}
    For any $\delta \in \left[0,1 \right)$ and $\alpha\in [0,1]$, the distortion achieved by $\SimVeto$ on 
    $\alpha$-decisive
    instances where the prediction $\p$ has error $\error = n \d(\p,\o) / \SC(\o,d)$
    is at most
        $$\frac{2 + \alpha -\alpha \delta}{1+\delta} + \min \left\{\frac{2\delta \error}{1+\delta},~ \frac{8\delta}{(1+\delta)(1-\delta)}\right\}.$$
\end{proposition}
\begin{proof}
For any $\delta \in [0,1)$, 
assume that $a$ is the candidate output by a given execution of $\SimVeto$ on an $\alpha$-decisive instance
when provided with a prediction $\p$ with error $\error=n d(\p, \o) / \SC(\o,d)$.
Proposition \ref{prop:robustness_mainalpha} already show that the worst-case distortion of $\SimVeto$ is at most $\left(\frac{2 + \alpha -\alpha \delta}{1-\delta} + \frac{8\delta}{(1+\delta)(1-\delta)}\right)$ for any $\delta$, so we just need to prove that it is also at most $\frac{2+\alpha-\alpha\delta+2\delta \error}{1+\delta}$.
Note that if $\p = \o$ then $\eta = 0$ and the claim follows from Proposition \ref{prop:consistencyalpha}.  We thus assume for the rest of the proof that $\p \neq \o$.

The social cost of $a$ can be upper bounded as follows:
\allowdisplaybreaks
\begin{align*}
    \SC(a,d) &= \sum_{v \in \V}d(v,a)\\
        & = \sum_{v \in \V} n \cdot \unip(v) \cdot d(v,a) \tag{by Def.\  of $\unip$}\\
        & = n \sum_{v \in \V} \sum_{c: (v,c) \in E_a} w((v,c)) \cdot d(v,a) \tag{by Def.\  \ref{def:fractionalperfectmatching}}\\
        & \leq n \sum_{v \in \V} \sum_{c: (v,c) \in E_a} w((v,c)) \cdot d(v,c) \tag{by Def.\  \ref{def:pqgraph}}\\
        & \leq n \sum_{v \in \V} \sum_{c: (v,c) \in E_a} w((v,c)) \cdot \left(d(v,\o) + d(\o, c)\right) \tag{triangle inequality}\\
        & \leq \SC(\o,d) + n \sum_{v \in \V} \sum_{c: (v,c) \in E_a} w((v,c)) \cdot d(\o, c) \\
        & = \SC(\o,d) + n \sum_{c\in \C} \sum_{v \in \V: (v,c) \in E_a} w((v,c)) \cdot d(\o,c)\\
        & =   \SC(\o,d) + n \sum_{c\in \C} \hat{q}(c) \cdot d(\o,c) \tag{by Def.\  \ref{def:fractionalperfectmatching}}\\
        & =  \SC(\o,d) + n \cdot \frac{1-\delta}{1+\delta}  \left(\sum_{c \neq \p} \frac{\plu(c)}{n} \cdot d(\o,c) +  \frac{\plu(\p)+\frac{2\delta n }{1-\delta}}{n} \cdot d(\o,\p)\right)\tag{by Def. \ref{def:algorithm}}\\
        & = \SC(\o,d)  +\frac{2\delta \error}{1+\delta}\SC(\o,d) + \frac{1-\delta}{1+\delta} \sum_{c \neq \o} \plu(c) \cdot d(\o,c) \tag{by $d(\o,\p)= \error \cdot \SC(\o,d)/n$}\\
        & = \frac{1+\delta + 2\delta \error}{1+\delta}\SC(\o,d) +\frac{1-\delta}{1+\delta}  \sum_{c\in \C} \sum_{v: \topc(v) = c} d(\o, c)\\
        & \leq \frac{1+\delta + 2\delta \error}{1+\delta}\SC(\o,d) + \frac{1-\delta}{1+\delta} \sum_{c\in \C} \sum_{v: \topc(v) = c} (d(c,v) + d(\o,v)) \tag{triangle inequality}\\
         & \leq \frac{1+\delta + 2\delta \error}{1+\delta}\SC(\o,d) + \frac{1-\delta}{1+\delta} \sum_{c\in \C} \sum_{v: \topc(v) = c} (\alpha d(\o,v) + d(\o,v)) \tag{$\alpha$-decisiveness}\\
        & \leq  \frac{1+\delta + 2\delta \error}{1+\delta}\SC(\o,d) + \frac{1-\delta}{1+\delta} \sum_{c}\sum_{v: \topc(v) = c} (1+\alpha)d(\o,v) \tag{since $c= \topc(v) \pref_v \o$}\\
        & \leq \frac{1+\delta + 2\delta \error}{1+\delta}\SC(\o,d) + \frac{(1+\alpha)(1-\delta)}{1+\delta}\SC(\o,d)
    \leq \left(\frac{2+\alpha-\alpha\delta+2\delta \error}{1+\delta}\right)\SC(\o,d). \qedhere
\end{align*}
\end{proof}

\end{document}